%% file: paper.tex
\documentclass[acmsmall]{acmart}
\setcopyright{none} 
\renewcommand\footnotetextcopyrightpermission[1]{} 
\AtBeginDocument{%
  \providecommand\BibTeX{{%
    \normalfont B\kern-0.5em{\scshape i\kern-0.25em b}\kern-0.8em\TeX}}}

\usepackage[utf8]{inputenc}
\usepackage{graphicx} 
\usepackage{svg} 
\usepackage{subfig} 
\usepackage{booktabs} 
\usepackage{adjustbox} 
\usepackage{multirow} 
\usepackage{array}
\usepackage{makecell}
\usepackage{amsmath} 
\usepackage{titlesec} %

\newcommand{\KL}[1]{{#1}}

\newcommand{\baijun}[1]{{#1}}

\titleformat{\paragraph}[hang]{\normalfont\normalsize\bfseries}{\theparagraph}{1em}{}
\titlespacing*{\paragraph}{0pt}{3.25ex plus 1ex minus .2ex}{1em}

\begin{document}
\pagestyle{empty}
\title{Beyond Fidelity: Explaining Vulnerability Localization of Learning-based Detectors}



\author{Baijun Cheng}
\authornote{Equal contribution}
\email{prophecheng@stu.pku.edu.cn}
\affiliation{
	\institution{Peking University}
    \streetaddress{5 Yiheyuan Road, Haidian District}
    \city{Beijing}
    \country{China}
    \postcode{100871}
 }

\author{Shengming Zhao}
\email{shengmi1@ualberta.ca}
\authornotemark[1]
\affiliation{
	\institution{University of Alberta}
    \streetaddress{11011-88 Avenue}
    \city{Edmonton}
    \state{Alberta}
	\country{Canada}
    \postcode{T6G 2G5}
 }
 
\author{Kailong Wang}
\authornote{Corresponding authors}
\email{wangkl@hust.edu.cn}
\affiliation{
	\institution{Huazhong University of Science and Technology}
    \streetaddress{1037 Luoyu Road, Hongshan District}
    \city{wuhan}
    \state{hubei}
	\country{China}
    \postcode{430074}
 }

\author{Meizhen Wang}
\email{mzwang@hust.edu.cn}
\affiliation{
	\institution{Huazhong University of Science and Technology}
    \streetaddress{1037 Luoyu Road, Hongshan District}
    \city{wuhan}
    \state{hubei}
	\country{China}
    \postcode{430074}
 }

\author{Guangdong Bai}
\email{g.bai@uq.edu.au}
\affiliation{
	\institution{University of Queensland}
    \state{St Lucia QLD}
    \city{Brisbane}
    \state{Queensland}
	\country{Australia}
    \postcode{4072}
 }

\author{Ruitao Feng}
\email{ruitao.feng@unsw.edu.au}
\affiliation{
	\institution{University of New South Wales}
    \streetaddress{High Street}
    \city{Sydney}
    \state{New South Whales}
	\country{Australia}
    \postcode{2052}
 }

\author{Yao Guo}
\email{yaoguo@pku.edu.cn}
\authornotemark[2]
\affiliation{
	\institution{Peking University}
    \streetaddress{5 Yiheyuan Road, Haidian District}
    \city{Beijing}
    \country{China}
    \postcode{100871}
 }

\author{Lei Ma}
\email{ma.lei@acm.org}
\affiliation{
	\institution{University of Alberta}
    \streetaddress{11011-88 Avenue}
    \city{Edmonton}
    \state{Alberta}
	\country{Canada}
    \postcode{T6G 2G5}
 }

\author{Haoyu Wang}
\email{haoyuwang@hust.edu.cn}
\authornotemark[2]
\affiliation{
	\institution{Huazhong University of Science and Technology}
    \streetaddress{1037 Luoyu Road, Hongshan District}
    \city{wuhan}
    \state{hubei}
	\country{China}
    \postcode{430074}
 }
 

\begin{abstract}
\textbf{Abstract}: Vulnerability detectors based on deep learning~(DL) models have proven their effectiveness in recent years.
However, the shroud of opacity surrounding the decision-making process of these detectors makes it difficult for security analysts to comprehend. 
To address this, various explanation approaches have been proposed to explain the predictions by highlighting important features, which have been demonstrated effective in domains such as computer vision and natural language processing. 
Unfortunately, there is still a lack of in-depth evaluation of vulnerability-critical features, such as fine-grained vulnerability-related code lines, learned and understood by these explanation approaches.
In this study, we first evaluate the performance of ten explanation approaches for vulnerability detectors based on graph and sequence representations, measured by two quantitative metrics including fidelity and vulnerability line coverage rate.
Our results show that fidelity alone is insufficient for evaluating these approaches, as fidelity incurs significant fluctuations across different datasets and detectors. \KL{We subsequently check the precision of the vulnerability-related code lines reported by the explanation approaches and find poor accuracy in this task among all of them.}
This can be attributed to the inefficiency of explainers in selecting important features and the presence of irrelevant artifacts learned by DL-based detectors. 
\end{abstract}



\received{9 July 2023}
\received[revised]{25 September 2023}
\received[accepted]{3 January 2024}

\maketitle
\renewcommand{\shortauthors}{Baijun Cheng et al.}

\input{introduction}
\input{background}

\input{motivation}

\input{studyDesign}

\input{experiments}
\input{relatedAndConclude}

\begin{acks}
This work was partially supported by the National Natural Science Foundation of China (No. 62141208), the National Natural Science Foundation of China (grant No.62072046), the Key R\&D Program of Hubei Province (2023BAB017, 2023BAB079), HUST CSE-HongXin Joint Institute for Cyber Security, and HUST CSE-FiberHome Joint Institute for Cyber Security, and the Cyber Security Agency of Singapore under its National Cybersecurity R\&D Programme (NCRP25-P04-TAICeN). 
Any opinions, findings, conclusions, or recommendations expressed in this material are those of the author(s) and do not reflect the views of the respective funding agencies.
\end{acks}

\bibliographystyle{unsrt}
\bibliography{references}

\end{document}

%% file: introduction.tex
\section{Introduction}
The ubiquity of software in today's digital society underpins the paramount importance of vulnerability detection techniques. As we propel further into the digital age, software permeates every corner of our lives, from healthcare to finance, transportation to communication. This pervasive deployment has led to an exponential increase in the surface area for potential attacks, amplifying the demand for robust, accurate, and efficient vulnerability detection mechanisms.
Current vulnerability detection techniques~\cite{Coverity, RATS, Infer, CSA, Fortify, Checkmarx, SVF,devign, reveal, ivdetect, dwk, SySeVR, vdp, vulcnn, Russel}, however, are often characterized by significant limitations that stymie their effectiveness. A core shortcoming lies in their limited ability to pinpoint the exact code lines that instigate vulnerabilities. Existing detection tools frequently flag entire segments of code as potentially hazardous, leading to vast sections needing review. This wide-net approach not only overwhelms developers with high rates of false positives but also sows the seeds of inefficiency and can potentially delay mitigation efforts. The ability to identify the precise lines of code responsible for a vulnerability, thus, is an issue of critical importance and poses an intricate challenge in the realm of program analysis.

In addressing the challenge, researchers have turned to the promising realm of explainable AI techniques. These methods illuminate the defining features within detected vulnerabilities, thereby equipping developers with the insights necessary to comprehend and rectify these issues more effectively. The essence of these techniques is in highlighting the key characteristics of vulnerable programs, playing a critical role in localizing vulnerability-correlated code statements. 
Such explanation techniques have been successfully integrated with Deep Learning (DL)-based detectors in the vulnerability detection area. For instance, Li et al.~\cite{ivdetect} incorporated the GNNExplainer~\cite{gnnexplainer} into their proposed detector, while Wu et al.~\cite{vulcnn} employed Grad-CAM++~\cite{gradcam++} in their image-based vulnerability detector, VulCNN. In another innovative approach, Zou et al.~\cite{IDLP} designed a heuristic search algorithm to identify pivotal tokens within code gadgets, generated by VulDeepecker~\cite{vdp} and SySeVR~\cite{SySeVR}. It is noteworthy that these explanation techniques are not confined to the field of vulnerability detection; they have found valuable applications in a variety of domains, including computer vision (CV)\baijun{~\cite{abhishek2022attribution}} and natural language processing (NLP)\baijun{~\cite{madsen2022post}}, primarily in the context of DL models used for vulnerability detection.

Despite the evident potential of existing explanation techniques, our observations have revealed a disconnect between the explanation results and their intrinsic relevance to the identified vulnerabilities. In some instances, the explainer models mistakenly allocate high relevance scores to code statements that bear no relation to the underlying vulnerabilities. This observation incited us to undertake a thorough performance analysis of these explanation techniques within the specific scope of vulnerability detection.

In our study, \textit{the overarching objective is to conduct a rigorous evaluation and exmaination of the performance of these explanation techniques within the field of fine-grained vulnerability localization.} We set our sights on evaluating five graph-based explanation methodologies: GNNExplainer~\cite{gnnexplainer}, PGExplainer~\cite{pgexplainer}, GNN-LRP~\cite{gnnlrp}, GradCAM~\cite{grad}, and DeepLift~\cite{deeplift}. 
These techniques are examined in conjunction with four graph-based vulnerability detectors, namely DeepWuKong~\cite{dwk}, IVDetect~\cite{ivdetect}, Reveal~\cite{reveal}, and Devign~\cite{devign}. Furthermore, we venture into the investigation of five sequence-based techniques—LRP~\cite{LRP-LSTM}, Shap~\cite{SHAP}, Lime~\cite{LIME}, GradInput~\cite{grad}, and DeepLift~\cite{deeplift}—in relation to sequence-based vulnerability detectors such as VulDeePecker~\cite{vdp}, SySeVR~\cite{SySeVR}, and TokenLSTM. 

In our evaluation process, we initially use fidelity as the metric to evaluate existing explanation techniques, a common practice in explainable AI to assess the accuracy of generated explanations. However, we find that relying solely on fidelity leads to significant fluctuations in scores across different detectors and datasets, showing inconsistent results. To address this, we introduce an additional metric called Vulnerability Line Coverage rate~(LC), mirroring MSR~\cite{ContraFlow}, which can be divided into two sub-metrics: the Vulnerability-Triggering Code Line Coverage rate~(TLC) and the Vulnerability-Fixing Code Line Coverage rate~(FLC). These aim to provide a more detailed understanding of how well the explanation techniques can identify and explain vulnerabilities. However, our findings show that most explanation techniques struggle to pinpoint key vulnerability-related statements accurately, as seen in both TLC and FLC measurements~(RQ2, \S~\ref{LC}). For instance, when applying GNNExplainer to IVDetect on the SARD dataset, the results show a wide range of TLC values from 7\% to 64\%, and FLC values from 18\% to 66\%, despite the fidelity scores remaining relatively high between 74\% and 92\%. This highlights that existing approaches, even with high fidelity scores, fail to learn and reflect the critical features inherent to the vulnerabilities, indicating that a broader evaluation beyond fidelity is necessary to truly understand the effectiveness of explanation techniques in addressing vulnerabilities.


Throughout our study, we have gained two in-depth understandings regarding the DL-based vulnerability detectors and explanation approaches, respectively. 
On the one hand, the existing explanation approaches are prone to pick up vulnerability-irrelevant features as explanations.
On the other hand, DL-based detectors indeed perform vulnerability detection by diligently learning the distinctions between vulnerable and normal samples. 
The varying sensitivities of model-related explainers on retrained and previous models to vulnerability-fixing lines suggest that they may primarily identify feature discrepancies between positive and negative samples, rather than understanding how vulnerabilities are triggered~(RQ 3.1, \S~\ref{sec:cmp}). 
To further validate our point and evaluate their comprehension of vulnerabilities, we manually patch a subset of vulnerabilities, using the originally patched samples as a reference. 
By comparing detection accuracy between non-vulnerability samples, we anticipate minimal accuracy differences if the detectors truly understand vulnerability nuances~(RQ 3.2, \S~\ref{sec:turnover}). 
Surprisingly, a significant portion of manually fixed codes is misclassified as vulnerable, with fewer false alarms in the originally patched samples. 
Moreover, we identify a kind of irrelevant text features that notably influence the model's predictions, collectively indicating that current detectors have not fully understood vulnerability semantics.

Our main contributions could be summarized as follows:

\begin{itemize}

\item \textbf{Enhancing the evaluation metrics for explanation approaches.} We employ the quantitative metric of fidelity and introduce additional metrics, namely TLC and FLC, to evaluate the efficacy of the explanation approaches in accurately localizing vulnerabilities. 
Our investigation reveals that none of the explanation approaches effectively localize vulnerabilities even when they achieve high fidelity scores. 

\item \textbf{Proposing a new evaluation metric for DL-based detectors.} To complement the existing evaluation metrics, we introduce a new metric VUR that is focused on the impact of vulnerability-unrelated perturbations on DL-based detectors.
The metric is employed to determine the proportion of perturbations unrelated to vulnerabilities. 
We observe that all detectors are affected to varying extents by vulnerability-unrelated perturbations, potentially causing a flip in their prediction outcomes.

\item \textbf{Revealing the limitations in DL-based detectors.} As deep-learning vulnerability detectors aim to identify vulnerabilities by learning distinctions between vulnerable and normal samples, we find that the models can still be misdirected by irrelevant disturbances. Furthermore, these models are unable to recognize vulnerability-fixing patterns not present in the training dataset. We also identify a specific text feature that can trigger false positives, indicating DL-based detectors primarily distinguish between vulnerable and normal samples without a full comprehension of the mechanisms underlying vulnerabilities.
\end{itemize}

To facilitate the reproductivity of our analysis, we open-source our codes and datasets. The website is https://github.com/for-just-we/VulDetectArtifact.

%% file: background.tex
\section{Background}\label{sec:overview}

In this section, we primarily introduce the DL-based detectors and explanation approaches used in this study.

\subsection{DL-based Vulnerability Detectors}

\baijun{In the realm of vulnerability detection, DL-based approaches have emerged as potent tools for identifying security vulnerabilities within software programs. These methodologies typically entail segmenting source code into discrete code fragments, subsequently vectorizing each fragment using deep learning models, and predicting the likelihood of a vulnerability existing within that fragment. The inherent advantage of these methods lies in their capacity to autonomously discern patterns and regularities within source code, thereby enhancing the accuracy and efficiency of vulnerability detection.
}

Based on how source code is segmented, DL-based vulnerability detectors can be categorized into two primary classes: function-level and slice-level detectors. In function-level detectors~\cite{Russel, funded, reveal, ivdetect, devign, linevd, linevul}, individual functions within the program are treated as isolated code fragments. In contrast, slice-level detectors~\cite{vulspg, dwk, vdp, SySeVR, vuldeelocator} extract code fragments spanning multiple functions based on data dependencies or control flow dependencies of vulnerability-critical statements. This approach offers a comprehensive contextual view of vulnerabilities, thus contributing to improved detection precision.

Furthermore, concerning the vectorization of code fragments, these vulnerability detectors can be further classified as sequence-based and graph-based detectors. Sequence-based methods directly parse source code fragments into token sequences and employ sequence neural networks for vectorization. 
Conversely, graph-based approaches typically parse the structures of code fragments, such as abstract syntax trees (AST), control flow graphs (CFG), or program dependency graphs (PDG), and further process these graph structures using graph neural networks. This facilitates the capture of structural information and dependency relationships within the source code, thereby enhancing the effectiveness of vulnerability detection.

Our study involves seven DL-based detectors: DeepWuKong~\cite{dwk}, Reveal~\cite{reveal}, Devign~\cite{devign}, IVDetect~\cite{ivdetect}, TokenLSTM, VulDeePecker~\cite{vdp}, SySeVR~\cite{SySeVR}. 
Table~\ref{tab:overview} provides an overview of these detectors and their key characteristics. Specifically:

\begin{itemize}
    \item \baijun{Devign extracts a comprehensive joint graph that encompasses the AST, CFG, Data Dependence Graph~(DDG), and Natural Code Sequence~(NCS) for each function.
    It generates initial embeddings for each node with Word2Vec~\cite{word2vec}.
    Also, its deep learning model is composed of a gated graph recurrent layer~(GGRN)~\cite{ggnn}, a convolutional layer, and an MLP Layer.}

    \item \baijun{Reveal follows a similar pipeline as Devign. 
    However, in comparison to Devign, Reveal represents each function using a code property graph~(CPG)~\cite{CPG}, which is constructed from the AST, CFG, and Program Dependence Graph~(PDG). 
    Additionally, when aggregating all node embeddings into a graph embedding, Reveal employs a summation approach for all node embeddings instead of using a convolutional layer.
    Reveal employs a similar pipeline with Devign.}

    \item \baijun{IVDetect extracts PDGs for each function and generates initial embeddings using sub-token sequences, variables, and types, as well as their surrounding contexts, by leveraging Glove~\cite{glove} and GRU~\cite{bgru}. 
    Subsequently, it employs a graph convolutional network~(GCN) model~\cite{gcn} to update node embeddings and aggregates them using a summation approach.
    To identify vulnerabilities at the statement-level, IVDetect selects crucial sub-graphs within each PDG using the GNNExplainer.~\cite{gnnexplainer}.}

    \item \baijun{DeepWuKong, like IVDetect, extracts PDGs to identify vulnerabilities. 
    However, it differs in its approach. DeepWuKong begins by inter-procedurally slicing a program, specifically focusing on statements that could potentially trigger vulnerabilities. 
    Subsequently, it embeds each sliced PDG using a combination of Doc2Vec~\cite{doc2vec} and GCN.}

    \item \baijun{TokenLSTM straightforwardly parses each function into a token sequence and utilizes a Bi-LSTM~\cite{blstm} to determine if the function might contain vulnerabilities.}

    \item \baijun{VulDeePecker initiates the detection process by first identifying statements that involve vulnerability-related system API calls and subsequently slices the program accordingly using DDG. 
    It then proceeds to vectorize and make predictions for each slice using Word2Vec~\cite{word2vec} and Bi-LSTM~\cite{blstm}.}

    \item \baijun{SySeVR expands upon VulDeePecker's approach by broadening the set of vulnerability-related statements and employing PDGs for slicing. 
    Additionally, it substitutes the use of Bi-LSTM with Bi-GRU~\cite{bgru} for the purpose of vulnerability detection.}
\end{itemize}

All of these DL-based detectors approach vulnerability detection as a binary classification problem, aiming to determine whether a given code fragment contains vulnerabilities. 
Initially, the code fragment is processed into a feature vector, which is then fed into a classifier to predict the probability of the code fragment containing vulnerabilities. 
It is worth mentioning that the TokenLSTM method is an implementation based on the detector proposed by Russel et al.~\cite{Russel}.  
To ensure consistency in our experiment settings, we implement the TokenLSTM method utilizing the bidirectional LSTM~(BLSTM)~\cite{blstm} model as the feature extraction layer, while excluding the random forest classifier~\cite{RF} from the architecture.


\begin{table}[t]
    \caption{Overview of seven DL-based detectors.}
    \centering
    \begin{adjustbox}{width=1.0\textwidth}
        \begin{tabular}{c|c|c|c}
        \hline
            Approaches & Representation & Granularity & Model \\ \hline
            
            TokenLSTM & token sequence~\cite{BTP} & function~\cite{BTP} & BLSTM~\cite{blstm} \\ \hline
            
            VulDeePecker~\cite{vdp} & token sequence & slicing~\cite{BTP} + DDG~\cite{CPG} & Word2Vec~\cite{word2vec} + BLSTM \\ \hline
            
            SySeVR~\cite{SySeVR} & token sequence & slicing + PDG~\cite{CPG} & Word2Vec + BGRU~\cite{bgru} \\ \hline
            
            DeepWuKong~\cite{dwk} & PDG & slicing + PDG & Doc2Vec~\cite{doc2vec} + GCN~\cite{gcn} \\ \hline
            
            Devign~\cite{dwk} & CFG~\cite{CPG} + DDG + AST + NCS & function & Word2Vec + GGRN~\cite{ggnn} \\ \hline
            
            Reveal~\cite{reveal} & CPG~\cite{CPG} & function & Word2Vec + GGRN \\ \hline
            
            IVDetect~\cite{ivdetect} & PDG & function & Glove + TreeLSTM~\cite{tree-lstm} + GCN \\ \hline
        \end{tabular}
    \end{adjustbox}
    \label{tab:overview}
\end{table}

\subsection{Explanation Approaches}

\baijun{
In recent years, there has been notable progress in the field of deep learning, particularly in its application to vulnerability discovery. However, it's important to recognize that these deep learning models typically excel in binary classification tasks. Despite their success, their decision-making processes remain largely concealed within black-box models, which can raise trust and transparency concerns.}

\baijun{This lack of transparency and interpretability has raised skepticism regarding their use in real-world security applications. Fortunately, explainability methods have emerged as a promising approach to address this issue. These methods offer a fresh perspective by shedding light on the inner workings of machine learning systems. It's essential to acknowledge that their implementation often requires considering the context outlined earlier.
}

We overview the explanation approaches employed in this study, categorized into the following two classes based on the type of target detectors.
Table~\ref{table:explainers} provides an overview of the explanation approaches employed in this study.
It is worth noting that in our research, the explanation approach DeepLift~\cite{deeplift} offers both graph-based and sequence-based versions, we consider them as distinct interpretation methods.
Specifically:

\begin{itemize}
\item \baijun{\textbf{LIME}~\cite{LIME} fits a surrogate model for every sample $x$ in the sample space to approximate the performance of target model $f(x)$, and the surrogate model is self-explainable. To fit a surrogate model, LIME samples a number of points through perturbing $x$, then preserves a set of perturbations ${x_1^{'},x_2^{'},...,x_n^{'}}$ near $x$ in the sample space. After that, LIME utilizes the target model to gain the label of each perturbation and uses them as the train set to fix a linear model, which has a similar decision boundary to the target model.}

\item \baijun{\textbf{SHAP}~\cite{SHAP} approximates the Shapley value of each feature in the sample concerning the final prediction. This approach is inspired by the foundation of LIME and involves redefining the loss function $\mathcal{L}$ and the weight kernel $\pi _x(x_i^{'})$ based on the Shapley value principle. The Shapley additive explanation values are computed through a regression process.}

\item \baijun{\textbf{Gradient-based}~\cite{grad} explanation methods quantify the impact of a feature, denoted as $x_i$, in the input sample on the model's output, $o$, by calculating the partial derivative of $o$ concerning feature $i$. In this research, we opt for $grad_{input}$, which represents a gradient-based method and is calculated as the dot product of the input vector and the partial gradient.}

\item \baijun{\textbf{LRP}~(i.e. layer-wise relevance propagation) represents another significant approach for leveraging the internal relationships within neural networks to estimate feature importance scores. In essence, LRP utilizes the backpropagation mechanism of neural networks to propagate the output scores (relevance) from the model's output layer all the way back to the input layer. Consequently, each input feature receives a relevance value, serving as an importance score. To apply the LRP method to LSTMs and GRUs, we employ the propagation rule introduced by Arras et al~\cite{LRP-LSTM}.}

\item \baijun{\textbf{DeepLift}~\cite{deeplift} amalgamates the principles of LRP and Integrated Gradients. Much like LRP, DeepLift utilizes a backward propagation process. However, it introduces the notion of reference input from Integrated Gradients. This inclusion helps mitigate discontinuities and yields enhanced performance across various tasks compared to standard LRP approaches. Notably, DeepLift has been extended to support graph models by Yuan et al~\cite{explain_survey}.}

\item \baijun{\textbf{GNNExplainer}~\cite{gnnexplainer} employs a dual approach for learning importance in the input data, incorporating soft masks for edges and hard masks for node features. These masks are initially initialized with random values and iteratively updated to maximize the mutual information between the original predictions and the new predictions. It's worth noting that, in this context, the node features mask is omitted due to the fact that each node is embedded with a low-dimensional word vector, and individual dimensions within the feature vector do not carry any specific semantic significance.}

\item \baijun{\textbf{PGExplainer}~\cite{pgexplainer} generates discrete masks for edges as explanation results. Differing from GNNExplainer, it employs a parameterized mask predictor to estimate the edge mask values. This predictor takes the embedding vectors of the two nodes linked by an edge as input and computes the associated mask value for that edge.}

\item \baijun{\textbf{Grad-CAM}~\cite{grad} is a widely adopted explanation model for image classifiers. It has been extended for use with graph models to assess the significance of various nodes. The fundamental concept revolves around the fusion of hidden feature maps and gradients to determine node importance.}

\item \baijun{\textbf{GNN-LRP}~\cite{gnnlrp} investigates the significance of various graph walks through score decomposition. This approach aligns well with deep graph neural networks, as these walks correspond to message flows during neighborhood information aggregation. GNN-LRP leverages high-order Taylor decomposition of model predictions to break down the ultimate predictions into distinct graph walks using the LRP technique.}
\end{itemize}

\begin{table}[htbp]
\caption{Basic information of explanation approaches.}
\begin{adjustbox}{width=0.9\textwidth,center}
\begin{tabular}{c|c|p{0.6\textwidth}}
    \hline
    Explanation Method & Representation & \makecell{Introduction}\\
    \hline
    GNNExplainer~\cite{gnnexplainer} & Graph & GNNExplainer trains mask for edges and node features to identify meaningful input information. These masks are updated to maximize the mutual information between original and updated predictions.\\
    \hline
    PGExplainer~\cite{pgexplainer} & Graph & PGExplainer trains a parameterized mask predictor to predict the edge mask value.
    The predictor takes embedding vectors of two nodes connected by an edge as input and produces a corresponding mask value for the edge.\\
    \hline
    GNN-LRP~\cite{gnnlrp} & Graph & GNN-LRP explores the importance of various graph walks through score decomposition.
    It decomposes model predictions into different graph walks using LRP\\
    \hline
    GradCAM~\cite{grad} & Graph & Pope et al.~\cite{grad} extend GradCAM~\cite{gradcam} to the realm of graph classification, where it initially calculates the gradients of the target prediction with respect to the final node embeddings.
    Subsequently, these gradients are averaged to derive the weights for each feature map.  \\
    \hline
    DeepLift~\cite{deeplift} & Both & DeepLift combines LRP and Integrate Gradient, utilizing their insights to enhance its backpropagation process. 
    By incorporating a reference input, DeepLift overcomes discontinuities and achieves superior performance in various tasks compared to traditional LRP methods. It is extended to support deep graph models by Yuan et al.~\cite{explain_survey}.\\
    \hline
    Lime~\cite{LIME} & Sequence & Lime employ a surrogate model to approximate the performance of the target model. By perturbing the input $x$, LIME samples neighboring points ${x_1^{'},x_2^{'},...,x_n^{'}}$ and obtains their labels through the target model.\\
    \hline
    Shap~\cite{SHAP} & Sequence & Shap estimates the Shapley values of each feature in the prediction by adjusting the loss function, weight kernel, and solving a regression problem, thus providing additive explanations.\\
    \hline
    Gradient * Input~\cite{grad} & Sequence & 
    Gradient * Input is an early-stage gradient-based attribution method that computes attribution by taking the partial derivative of the output with respect to the input and multiplying it by the input itself. \\
    \hline
    LRP~\cite{LRP-LSTM} & Sequence & LRP is a technique that approximates feature importance scores by leveraging the internal relationship between input and output in neural networks. It uses backpropagation to propagate relevance scores from the output layer to the input layer, assigning importance values to each input feature.\\
    \hline

\end{tabular}
\end{adjustbox}
\label{table:explainers}
\vspace{0.1in}
\end{table}

%% file: motivation.tex
\section{Motivating Example}
\label{sec:motivation}

\begin{figure}[t]
  \centering\includegraphics[width=1.0\textwidth]{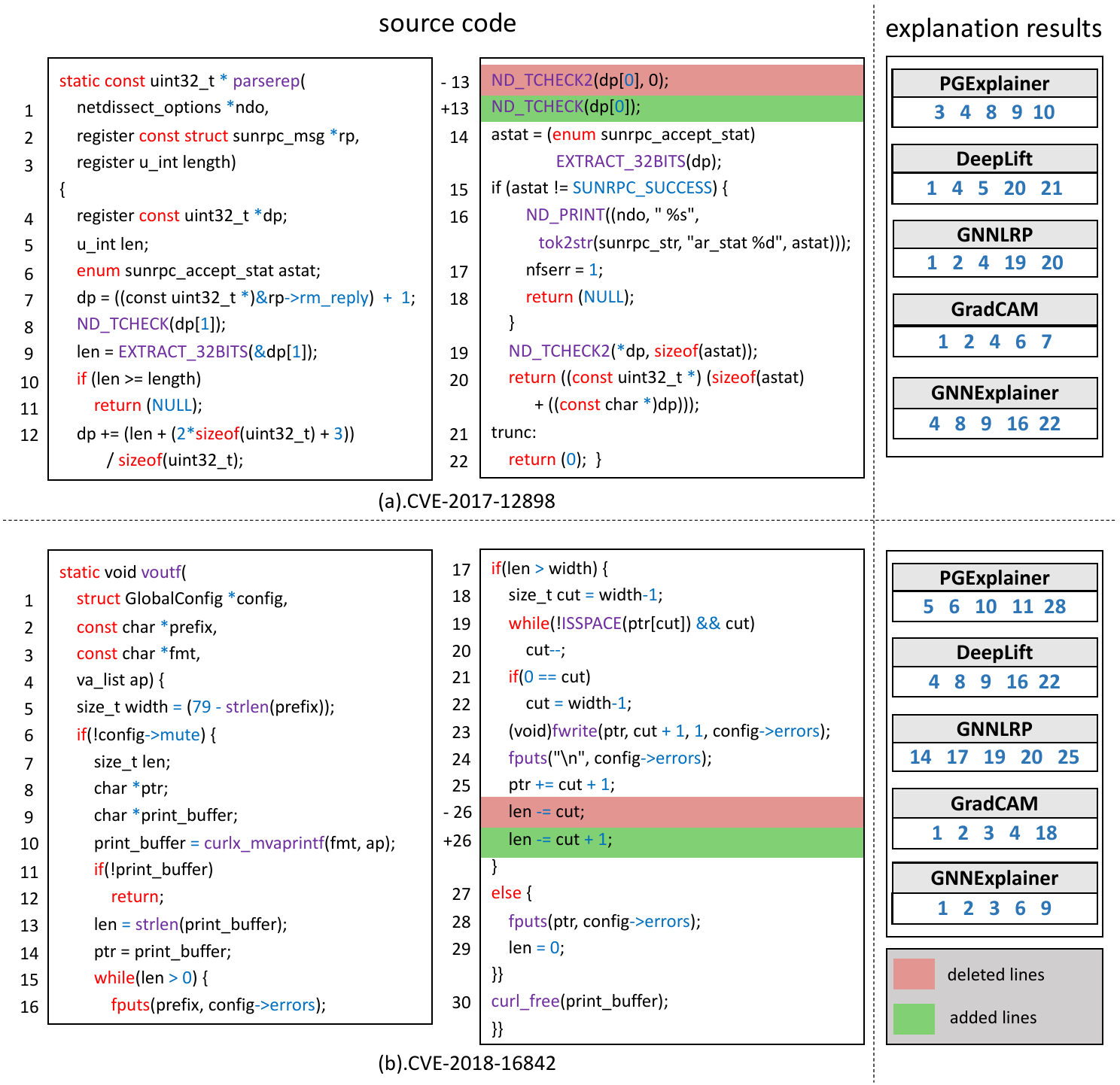}
  \caption{\baijun{An illustrative example featured by CVE-2017-12898 and CVE-2018-16842. In CVE-2017-12898, a modification in line 13 can be observed after patching, and in CVE-2018-16842, line 26 can be seen to be modified.
  The detector Reveal successfully identifies these two vulnerable functions. However, when employing five graph-based explainers to ascertain root causes, none of them highlight statement 13 in CVE-2017-12898 or statement 26 in CVE-2018-16842 in connection to the vulnerability.}}
  \vspace{-3mm}
\label{fig:example}
\end{figure}

\baijun{
Fig.~\ref{fig:example} (a) presents the function ``parserep'' in the tcpdump project, designed for parsing reply packets. 
This code is identified as vulnerable in Common Vulnerabilities and Exposures~(CVE-2017-12898) listed in the National Vulnerability Database~(NVD). The associated commit log for the fix indicates that the modification addresses bounds checking for the NFSv3 WRITE procedure. 
It checks whether the byte count is present in the captured data and whether the length of the opaque data being written is within the captured data, furthest forward in the packet, not the item before it. Following the patch, the statement in line 13, ``ND\_TCHECK2(dp[0], 0);'', was replaced with ``ND\_TCHECK(dp[0]);''
}

\baijun{Furthermore, Fig.~\ref{fig:example}(b) exhibits the function ``voutf'' within the curl project, which is associated with CVE-2018-16842. 
The commit log for this fix clarifies that it corrects an issue involving improper arithmetic when outputting warnings to stderr. 
Specifically, we can observe that in line 26, ``len -= cut''has been replaced with ``len -= cut + 1;''.
}

\baijun{While a learning-based detector, Reveal~\cite{reveal}, successfully identifies these two vulnerable functions, our investigation reveals that when employing five distinct explanation methods to pinpoint the root cause of the vulnerability, none of them can identify line 13 in ``parserep'' function or line 26 in ``voutf'' function as being related to the vulnerabilities. 
Furthermore, these five explanation methods yield diverse results, underscoring the substantial variability in their explanations.
}

\baijun{This observation may raise questions for users, leading them to question the reliability of explanation outputs produced by explanation approaches as well as DL-based detectors. Given the limited availability of systematic studies on explanation approaches in the context of vulnerability detection, our goal is to conduct this evaluation study to address this gap in the existing literature.}

%% file: studyDesign.tex
\section{Study Design}\label{sec:design}
In this section, we first present our evaluation framework concerning the DL-based detectors and the explanation approaches.
Subsequently, we elaborate on the evaluation metrics employed for the explanation approaches in \S~\ref{subsec:metric}.
Lastly, we introduce our pipeline for dataset collection and preprocessing in \S~\ref{subsec:dataset}.

\subsection{Evaluation Framework}

Fig.~\ref{fig:evaluation_framework} illustrates our evaluation framework for the DL-based detectors and explanation approaches.

\subsubsection{Research Questions (RQ)}\hfill

We address the following research questions~(RQs) in this study:

\emph{\textbf{RQ1}: Can the explanation approaches accurately describe the behavior of the models?}

To answer this question, we evaluate the explainers by fidelity. A higher fidelity score indicates better performance of the explainers in descriptive accuracy~\cite{XAIdefinition}.

\emph{\textbf{RQ2}: Can the explanation approaches provide vulnerability-relevant explanations?}

To answer this question, we assess the explainers with TLC and FLC scores. A higher TLC/FLC score reveals the better performance of the explainers in locating vulnerability-triggering statements~(VTS)/vulnerability-fixing statements~(VFS).

\emph{\textbf{RQ3}: Reliability of DL-based detectors?} 

We further divide this RQ into two distinct sub-questions, including can explanation approaches perform better on retrained models~(RQ3.1)? And can those detectors learn vulnerability-triggering mechanisms~(RQ3.2)?

\subsubsection{Framework Design}\hfill

    
    
\begin{figure}[t]
  \vspace{2mm}
  \centering\includegraphics[width=\textwidth]{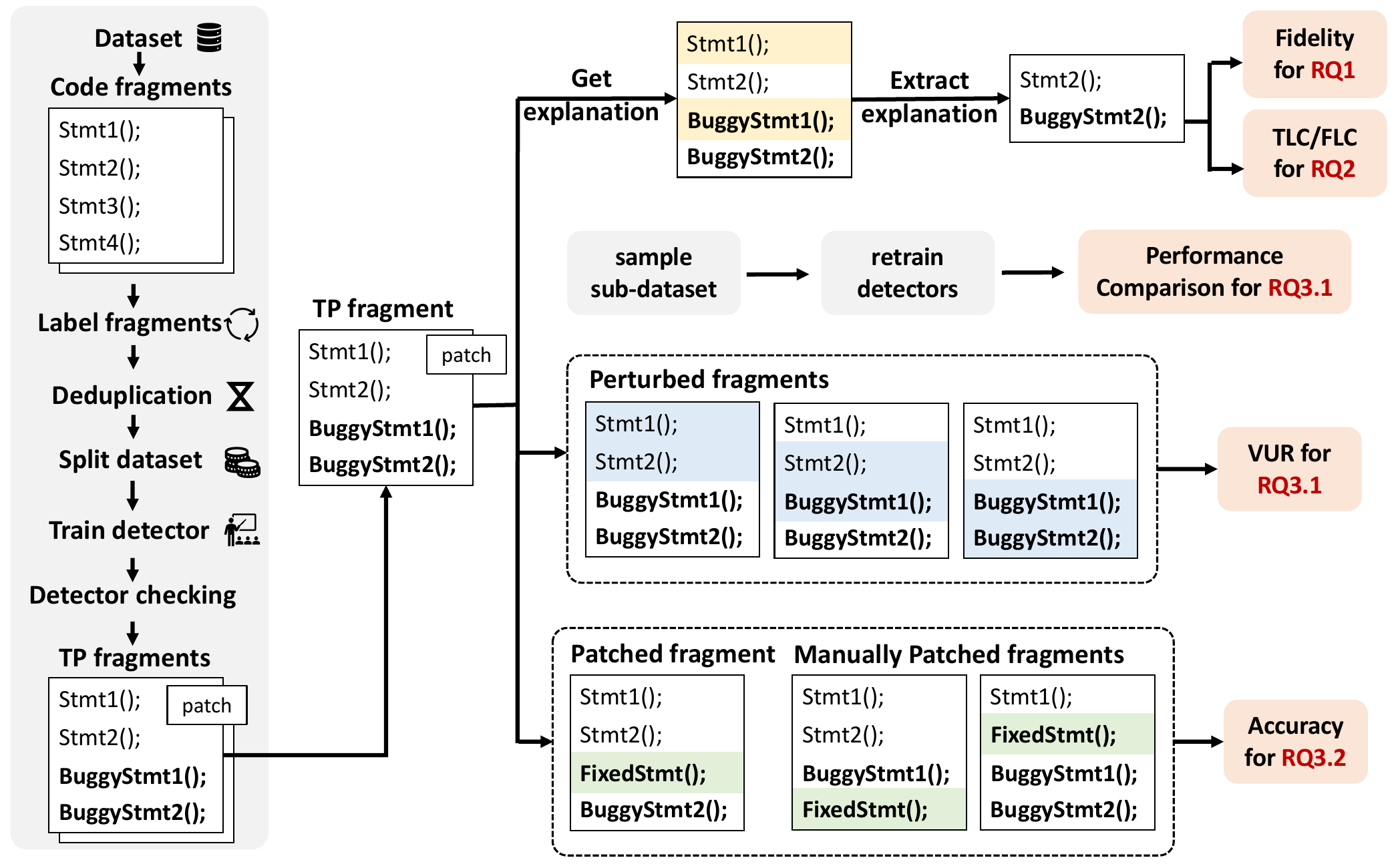}
  \caption{Evaluation framework.}
  \vspace{-2mm}
\label{fig:evaluation_framework}
\end{figure}

In this study, our objective is to explore the vulnerability localization effectiveness of explanation approaches combined with DL-based detectors, as illustrated by Fig.~\ref{fig:evaluation_framework}. We first extract code fragments from collected datasets and label them, perform deduplication and split them into train-set, test-set, and eval-set for further analysis. 
Then we follow previous work~\cite{dwk} to employ six widely-used metrics accuracy~(ACC), false positive rate~(FPR), false negative rate~(FNR), recall rate~(R), precision~(P), and F1 score~(F1) to evaluate the performance of the DL-based detectors in identifying vulnerable code fragments.
Through initial evaluation, we identify and select the true positive~(TP) samples that the detectors successfully detect from test-set.

To evaluate the explanation approaches, we first need to assess whether these explainers can identify important features for the detectors from the samples~(RQ1). 
Subsequently, we evaluate whether the important features selected by the explanation methods correspond to vulnerability-related statements~(RQ2). 
For this purpose, we conduct extensive experiments on TP samples from SARD~\cite{Sard} and Fan datasets~\cite{FanData}, obtaining fidelity scores and TLC/FLC scores corresponding to the explanation methods and perform further analysis.
We evaluate the explanation methods only on TP samples for two main reasons. 
First, non-vulnerable code segments are prevalent, rendering vulnerability localization on benign samples not necessary. Second, since the explanation methods operate on top of the detectors' predictions, vulnerability localization becomes irrelevant if the detector fails to recognize a particular vulnerable code snippet.

To explore the potential acquisition of features unrelated to vulnerabilities by DL-based detectors and its effect on the reliability of explanation results, we undertake two experiments~(RQ3).
In the initial segment~(RQ 3.1), we sample a balanced sub-dataset from the original full dataset where distinctions between vulnerable and normal samples are strictly constrained to the textual aspects of the VFS. 
Retraining a model on this subset while ensuring similar detection accuracy between the retrained and previous models enables us to compare the performance of the explainers. 
This approach aids in providing a deeper understanding of both the detectors and the explainers.
Additionally, we examine the susceptibility of the detectors' predictions to be altered by modifications unrelated to vulnerabilities by analyzing the ratio of vulnerability-unrelated perturbations (VUR) among all label-flipping perturbations.
In the subsequent segment~(RQ 3.2), we assess whether the detector comprehends vulnerability-triggering mechanisms by analyzing the disparities in detection accuracy between the original patched samples in the dataset and the manually repaired versions of those TP samples.

\subsection{Quantitative Metrics For Explanation Approaches}\label{subsec:metric}

Evaluating explanation methods is an ongoing question in research. 
To comprehensively evaluate these methods, we leverage insights from reference~\cite{XAIdefinition} and design metrics from two perspectives: \textbf{Descriptive Accuracy} and \textbf{Relevancy}. 
Descriptive Accuracy assesses if explanation methods accurately describe the behavior of models, while Relevancy determines if they provide valuable information for specific tasks. We approximate descriptive accuracy using fidelity~\cite{fidelity} and relevancy using LC.

\subsubsection{Fidelity}\hfill

Fidelity\cite{fidelity} measures how well interpretation methods capture important features identified by the target models. Higher fidelity indicates a better reflection of the models' behavior.
For sequence-based detectors, we mutate a set of $k$ important tokens identified by explanation methods to obtain modified instances. We then calculate the prediction scores of these modified instances in the target model and track the change in the average prediction score.
\begin{equation}
fidelity(g, f)=\frac{1}{T}\sum_i^T{f(x_i)-f(x_i^*)}
\end{equation}
In the above equation, $T$ represents the dataset length, $g(.)$ is the explanation method, and $f(.)$ is the target vulnerability detection model. $f(x_i)$ represents the probability of sample $x_i$ being classified as vulnerable by model $f(.)$. 
\baijun{$x_i^{*}$ denotes modified version of $x_i$.}

For graph-based detectors, the explanation results consist of a set of $k$ important nodes selected by explainers. We evaluate the explainers using the fidelity+ and fidelity- metrics, following the approach by Yuan et al.~\cite{explain_survey}.

\begin{equation}
    \begin{aligned}
        fidelity+(g, f) &= \frac{1}{T}\sum_i^T{f(g_i)-f(g_i^{1 - m_i})}\\
        fidelity-(g, f) &= \frac{1}{T}\sum_i^T{1- (f(g_i)-f(g_i^{m_i}))}
    \end{aligned}
\end{equation}

$g_i = G(X, E)$ represents the input graph data, where $X$ is the embedding matrix for nodes and $E$ is the edge set. The explanation is a discrete importance map $m_i$, with elements assigned 1 for critical nodes and 0 otherwise. The sum of $m_i$ indicates the maximum number of nodes in the manually set explanation subgraph. We modify the graph data by masking feature vectors of important nodes in $g_i^{1 - m_i} = G(X \cdot (1 - m_i), E)$ and retaining only the important features in $g_i^{m_i} = G(X \cdot m_i, E)$.

\subsubsection{LC}\hfill

The primary objective of employing the explanation approach in this context is to identify the decision basis for individual predictions and locate vulnerabilities at the statement level. If an explainer only generates a few carefully selected features that are unrelated to the vulnerability, they are of no value to explainer users. Therefore, LC is utilized to assess the correlation between the explanation results and vulnerabilities.
Formally, let $s^e$ represent the set of important tokens or nodes selected by explainers and $s^v$ denote the set of nodes or tokens present in labeled vulnerability statements. 
We calculate LC in the following equation:

\begin{equation}
    LC = \frac{|s^e \cap s^v|}{|s^v|}
\end{equation}


In this study, we examine LC from two perspectives.
To assess the ability of explainers to identify VTS, we employ TLC. In this context, $s^v$ represents the set of these triggering statements.
Furthermore, we utilize FLC to evaluate the capability of explainers in locating VFS. 
Here, $s^v$ represents the statements in the vulnerable code that have been modified in the fixed version.
Based on our observations, we note that VTS are present in both vulnerable and safe codes, whereas VFS differ between the two sample types.


\subsection{Dataset Construction}\label{subsec:dataset}

\subsubsection{Data Source}\hfill

The dataset used in this study must include annotations for lines of vulnerable code. However, certain real-world datasets like Devign and Reveal only label vulnerable functions~\cite{ivdetect}, which renders them unsuitable for our specific investigation. In the case of the D2A dataset~\cite{D2A}, it labels vulnerable traces in a program using the Infer tool ~\cite{Infer}. Although differential analysis on commit history is employed to reduce false positive rates, many cases still result in false positives. Croft et al.~\cite{dataQuality} also highlight the presence of over two-thirds of inaccurately labeled samples in D2A, primarily due to false positives in Infer warnings and mislabeled commits. Consequently, we have chosen not to include the D2A dataset in our study.

Therefore, we constructed our dataset by combining two sources: (1) the Software Assurance Reference Dataset~(SARD)~\cite{Sard}, a widely utilized vulnerability database, and (2) the Fan dataset~\cite{FanData}, which encompasses Common Vulnerabilities and Exposures~(CVEs) from 2002 to 2019 and provides 21 features for each vulnerability.

In the SARD dataset, each program or test case can be associated with one or more Common Weakness Enumeration (CWE) IDs, as a program can exhibit multiple vulnerability types. 
For our study, we follow Nie et al.~\cite{nie2023understanding} to primarily focus on six of the top 30 most critical C/C++ software weaknesses in 2021, specifically CWE20, CWE119, CWE125, CWE190, CWE400, and CWE787.
To acquire the relevant programs pertaining to these vulnerabilities, we utilize the same crawler employed in DeepWuKong~\cite{dwk}, which enables us to retrieve all the available programs associated with the aforementioned vulnerabilities.

To assess the performance of detectors and explainers on complex real-world open-source projects, we conducted additional studies using the Fan dataset. This dataset consists of over 10,000 vulnerable functions along with corresponding fixed code.
It is worth noting that Croft et al.~\cite{dataQuality} have also identified inaccuracies in the labeling of the Fan dataset, primarily attributed to difficulties in accurately tracing the vulnerability-related statements in fixing commits. 
These inaccuracies could potentially impact the experimental results and should be taken into consideration.

\subsubsection{Preprocessing of the Dataset}\hfill

Once the datasets have been downloaded, the initial step involves extracting code fragments at both the function level and slice level from these programs and subsequently labeling them.
We follow previous works~\cite{ivdetect, reveal, devign, dwk, vdp, SySeVR, Russel} for this step with tools Joern~\cite{Joern} and SVF~\cite{SVF}.

\begin{figure}[t]
  \centering
   \includegraphics[width=0.6\textwidth]{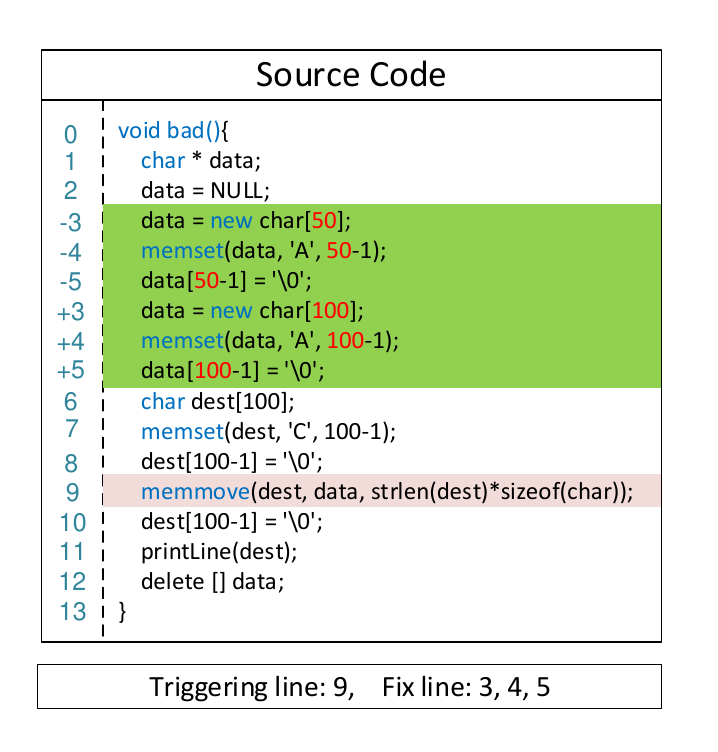}
  \vspace{-2mm}
  \caption{Illustration of a vulnerable code fragment and its corresponding fixed version with VTS and VFS labelled.}
  \label{fig:sardlabel}
\end{figure}

\textbf{\textit{Labeling SARD.}} Regarding the labeling of the SARD dataset, the VTS have already been labeled using XML files. 
After extracting all the code fragments from the original programs, we label each fragment as vulnerable if it contains lines that have been marked as vulnerable in the corresponding XML file, or vice versa. 
Simultaneously, the vulnerable lines in the vulnerable samples are labeled as \textbf{triggering lines}.
However, the VFS in the vulnerable samples of SARD are not labeled. 
Fig.~\ref{fig:sardlabel} provides an example of a vulnerable code snippet from the SARD dataset, along with its fixed version and labels. 
It is noteworthy that most vulnerabilities in SARD have a fixed repair mode. 
Taking this into account, we have implemented an automated labeling mechanism to identify and label the fixing line in the vulnerability code fragments.


\textbf{\textit{Labeling Fan.}} The Fan dataset is constructed by mining code change information from committed version patches to identify modified lines of code. It includes two types of modifications: added lines and deleted lines. We label the deleted lines in vulnerable functions and the lines in vulnerable functions with data or control dependencies by following the approach used in IVDetect~\cite{ivdetect}. 
The added lines in the patched version are considered fixed lines corresponding to the vulnerable lines. Notably, triggering lines are not labeled in Fan.


\begin{figure}[t]
  \centering
    \includegraphics[width=0.5\textwidth]{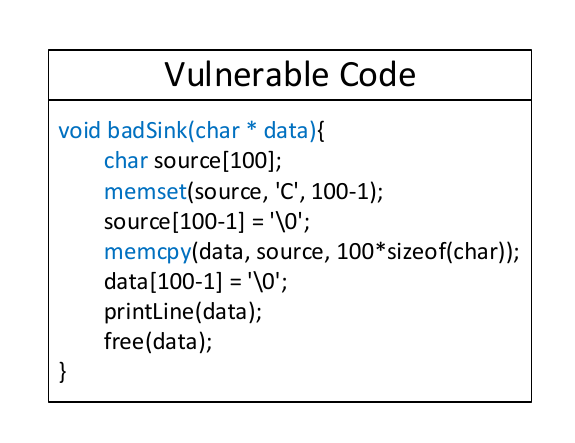}
  \vspace{-5mm}
  \caption{An example of a controversial code fragment.}
  \vspace{-5mm}
  \label{fig:mislabel}
\end{figure}

\textbf{\textit{Handling duplicate/conflict samples.}} Duplicates in the datasets can lead to overfitting and inflated performance metrics. 
To address this, we follow previous work~\cite{dwk} to employ a method based on MD5 value comparison to remove duplicate samples. 
Additionally, there may be samples labeled as both vulnerable and non-vulnerable due to mislabeling in the SARD and Fan datasets. 
Fig.~\ref{fig:mislabel} provides an example from the SARD dataset, where a sample is labeled as both vulnerable and non-vulnerable. 
These samples are typically referred to as benign in a vulnerable context within the ``main'' functions of SARD when labeled as vulnerable, and in a non-vulnerable context when labeled as non-vulnerable.
A similar situation occurs in the Fan dataset, where some code change information is extracted from consecutive patches. 
Consequently, a fixed code fragment may be modified again in the subsequent patch, resulting in certain functions being labeled as both vulnerable and non-vulnerable. In light of this observation, we label these conflicting samples as vulnerable.
In the last, we split each dataset into train-set, eval-set, and test-set with a ratio of 80\%-10\%-10\%.

%% file: experiments.tex
\section{Experiments}\label{sec:experiments}

\subsection{Experimental Setup}

The DL-based detectors are implemented following previous works~\cite{dwk, reveal, ivdetect, devign}. 
We utilize utilize PyTorch and Torch\_geometric for implementation of graph-based detectors, while employ Keras for sequence-based ones.
The explanation approaches are implemented following Yuan et al.~\cite{explain_survey}.
We first overview the detection performance of the DL-based detectors in \ref{subsec:VDA}.
Then in following parts, we present and analyze the experimental results for each RQ.


To facilitate the evaluation, we first derive a baseline by randomly selecting a certain number of nodes or tokens as explanation results for each sample and compute the corresponding fidelity and LC. Intuitively, the baseline gives an estimation of the fidelity and LC when a detector learns ``nothing'' relevant to the vulnerability. 
To reduce the influence of randomness, we repeat this process five times for each sample and calculate the average as the final result.
We denote this method as \textbf{\textit{Random}}.

For the explanation outputs, we present the top-k results, where k can take values of 1, 3, 5, and 7.
Top-k results refers to the k nodes or tokens with the highest weights for graph-based and sequence-based methods respectively.
Due to variations in the number of vulnerable lines across different datasets and the presence of multiple tokens in a labeled vulnerable line in the sequence-based approaches, the LC may not reach 100\% even if all top-k results match the vulnerable lines' nodes or tokens.
Considering this, we propose a benchmark value of TLC/FLC referred to as \textbf{\textit{Expectation}} to complement our evaluation. It is calculated by assuming all the top-k results are VTS/VFS, which can be intuitively interpreted as the upper bound of TLC/FLC.

For simplicity, we abbreviate DeepWuKong, Reveal, IVDetect, Devign, SySeVR, VulDeePecker, and TokenLSTM as DWK, Rev, IVD, Dev, SySe, VDP, and TL in the figures and tables.

\subsection{Vulnerability Detector Analysis}\label{subsec:VDA}

\begin{figure}[h!]
  \centering
  \includegraphics[width=1\textwidth]{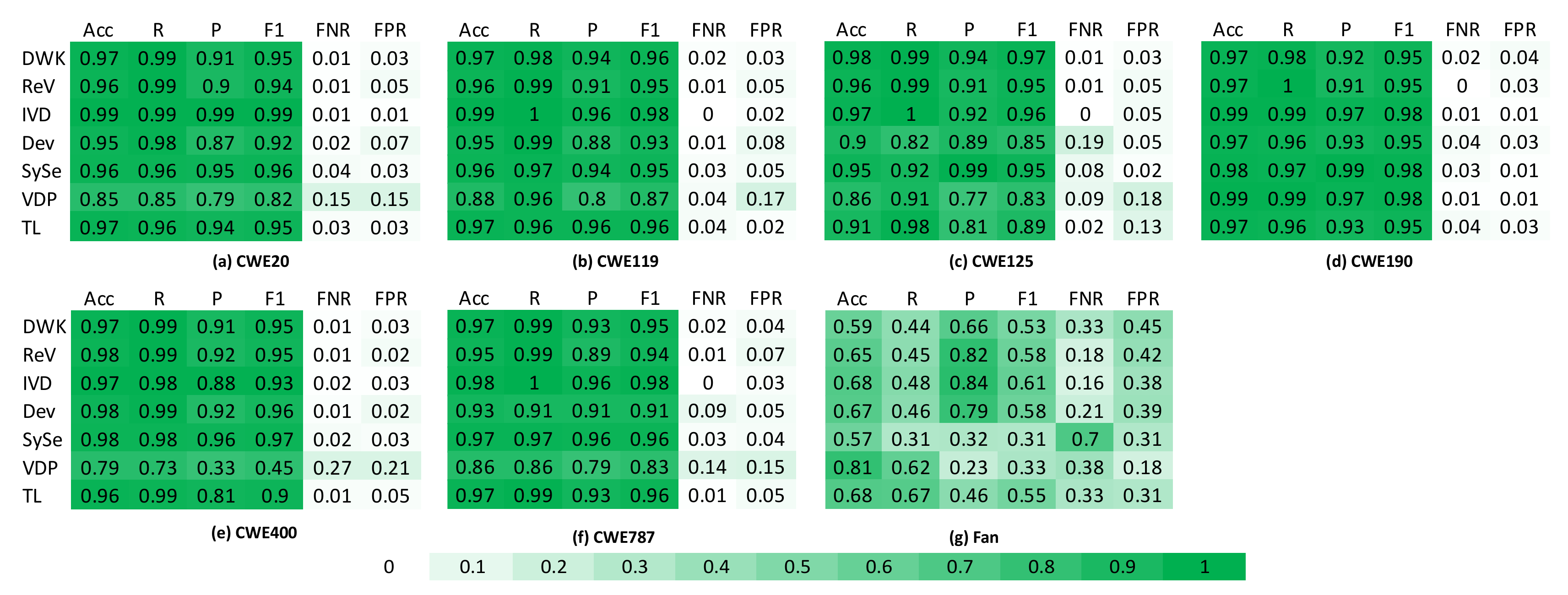}
    \caption{Detection performance of seven DL-based detectors in different datasets.
    \label{experiment:detection_result}}   
\end{figure}

In this study, we adopt the provided settings used in the literature or re-implement them for optimal F1 scores. We stop the training process if the F1 score shows no improvements for 10 iterations. 
The detection results of the DL-base detectors are shown in Fig.~\ref{experiment:detection_result}.

We observe that most detectors achieve high performance for the SARD dataset with F-measures exceeding 90\%. VulDeePecker shows relatively lower performance due to its limited consideration of data-dependence relations and overlooking conditional judgment statements that encompass vulnerable statements. We further observe that the results do not have a significant difference among six sub-datasets from SARD in the following experiments. Therefore, we comprehensively take all sub-datasets into consideration, and present the average results over the six datasets for simplicity in the subsequent evaluations~(RQ1-RQ3).

However, the performance of all detectors on the Fan dataset is significantly lower compared to that on SARD. This can be attributed to the greater complexity of code samples in real-world datasets. Nevertheless, it motivates us to investigate the features learned by these detectors. We initially hypothesized that the detectors may have learned specific text patterns unrelated to vulnerabilities that commonly appear in vulnerable code snippets.

\subsection{RQ1: Can the Explanation Approaches Accurately Describe the Behavior of the Models?}\label{subsec:DA} 

In this section, our objective is to assess the significance of the explanation outcomes on deep learning models.
To answer RQ1 we initially carry out individual experiments on graph-based and sequence-based detectors separately.

\subsubsection{Fidelity Among Graph-based Explainers}\hfill

\begin{figure}[h!]
  \vspace{-3mm}
  \includegraphics[width=1.0\textwidth]{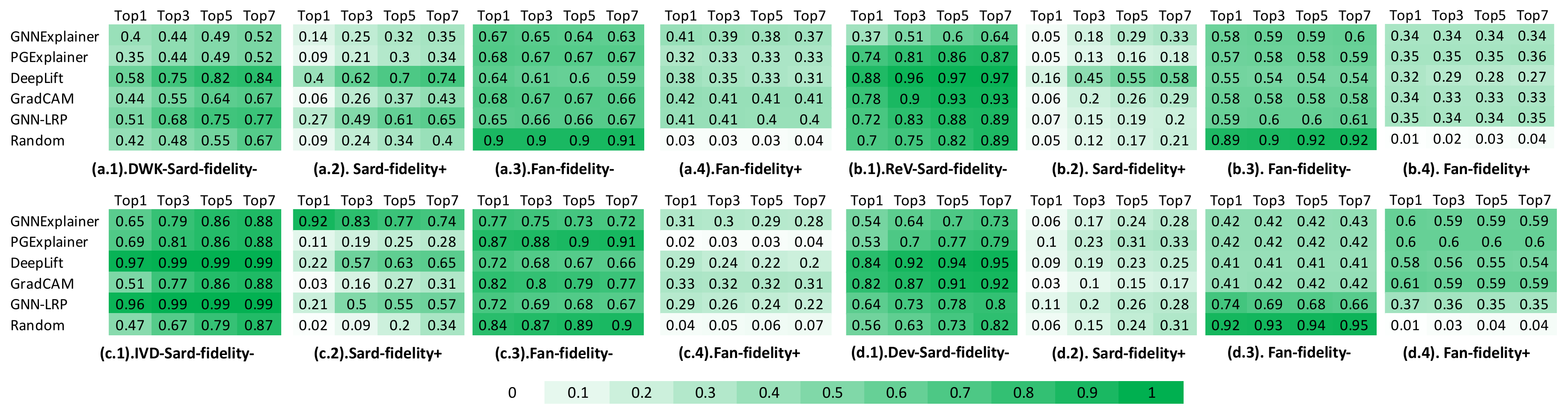}
    \vspace{-2mm}
    \caption{Mean fidelity scores of graph-based explainers on four detectors over six vulnerability types in SARD and performance on Fan datasets.
    \label{experiment:fidelity_graph}}   
\end{figure}

Fig.~\ref{experiment:fidelity_graph} shows the mean experiment results of the graph-based explanation models on four detectors for both SARD and Fan datasets. We observe that:

\par \textbf{\textit{The fidelity scores vary with calculation methods.}} We can observe that all explanation approaches, including Random, exhibit promising fidelity- performance. 
However, when interpreting the performance of the four graph-based detectors on Fan datasets, the Random shows high fidelity- but low fidelity+ scores. 
This indicates that random input perturbations minimally affect the model's prediction score, leading to artificially inflated fidelity- scores. Consequently, fidelity- may not be an adequate evaluation metric.
Hence in later parts, we mainly focus on fidelity+ results in the discussion due to its higher fluctuations observed.

\par \textbf{\textit{Fidelity scores of the same explanation method vary significantly across different datasets for a given detector.}} 
We can observe that the fidelity+ scores differ between Fan and SARD datasets.
This disparity is exemplified in Fig.~\ref{experiment:fidelity_graph} (c.2) and (c.4), which present the results of all five approaches applied to explain IVDetect.
Typically, explainers achieve higher fidelity scores by masking more nodes on SARD. However, this trend differs in the Fan dataset which can be illustrated in Fig.~\ref{experiment:fidelity_graph} (b.2) and (b.4) where DeepLift is employed to explain Reveal.

\par \textbf{\textit{The explainers exhibit varying fidelity scores when interpreting different detectors on identical datasets.}} For instance, as shown in Fig.~\ref{experiment:fidelity_graph} (b.2) and (d.2), on SARD, The fidelity+ scores of GradCAM on Devign are lower compared to those on Reveal. 
Similarly, GNNExplainer achieves the highest fidelity+ scores for IVDetect but lower for Devign as evidenced in Fig.~\ref{experiment:fidelity_graph} (c.2) and (d.2). This pattern can be observed in other cases as well.

\begin{figure}[t]
  \hspace{-3mm}
    \centering
    \includegraphics[width=1\textwidth]{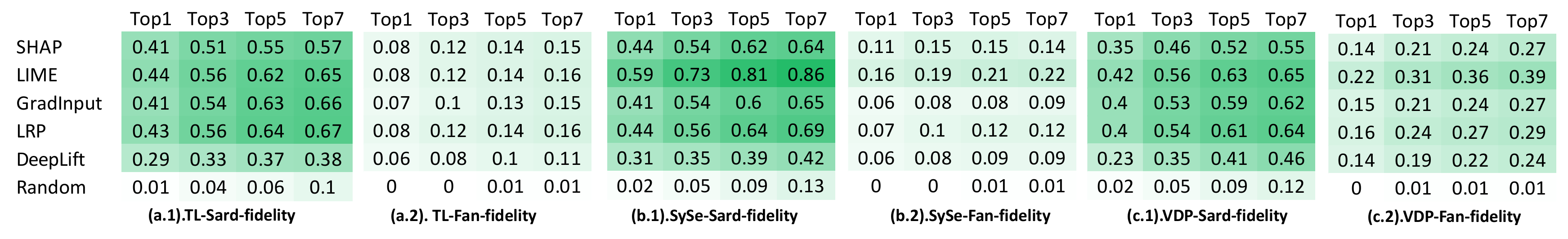}
  \vspace{-2mm}
  \caption{Mean fidelity scores of sequence-based explainers on three detectors over six vulnerability types in SARD, and performance on Fan dataset.}
\label{experiment:fidelity_seq}
\end{figure}

\subsubsection{Fidelity Among Sequence-based Explainers}\hfill

Fig.~\ref{experiment:fidelity_seq} illustrates the fidelity results of five sequence-based explainers for detectors TokenLSTM, SySeVR, and VulDeePecker. Our results bring the following observations: 

\par \textbf{\textit{The fidelity scores of the explainers continue to exhibit variability across different detectors and datasets. Leading to differing comparisons between them.}} 
On the Sard dataset, GradInput achieves the highest average fidelity scores on TokenLSTM, while Lime attains the highest scores on VulDeePecker and SySeVR.
While DeepLift respectively has the lowest performance for TokenLSTM, SySeVR and VulDeePecker as shown in Fig.~\ref{experiment:fidelity_seq} (a.1),(b.1),(c.1).
On the Fan dataset, Lime emerges as the superior explanation approach among the five with fidelity scores as illustrated in Fig.~\ref{experiment:fidelity_seq} (a.2),(b.2),(c.2).

\par \textbf{\textit{The explainers generally demonstrate comparatively higher fidelity scores than Random.}} We can observe the fidelity scores of those explanation approaches in both SARD and Fan are significantly higher than that of Random. 
For instance, when explaining SySeVR on SARD, the performance of Random is much lower than the results of the worst-performing explainer DeepLift as shown in Fig.~\ref{experiment:fidelity_seq} (b.1).

\vspace{2mm}
\noindent\fbox{
	\parbox{0.95\linewidth}{
ANSWER: The same explanation method exhibits varying performance across different detectors or datasets. 
It can achieve significantly better results than Random in some cases, while the opposite outcome can also be observed. 
As a result, determining the level of description accuracy attained by the explainer becomes challenging.
}}
\vspace{2mm}


\subsection{RQ2: Can the Explanation Approaches Provide Vulnerability-relevant Explanations?}\label{LC} 

In this section, our objective is to assess the relevance of the explanation outcomes to vulnerabilities, focusing primarily on two key statements: \textbf{VTS} and \textbf{VFS}.
To answer RQ2, we conducted experiments using two kinds of LC: TLC and FLC.  
Graph-based explainers aim to achieve the average recall rate of vulnerability-related statements, while sequence-based explainers focus on the average recall rate of tokens covered by vulnerability-related statements.

\subsubsection{TLC Among Graph-based Explainers}\hfill

\begin{figure}[h!]
  \includegraphics[width=1.0\textwidth]{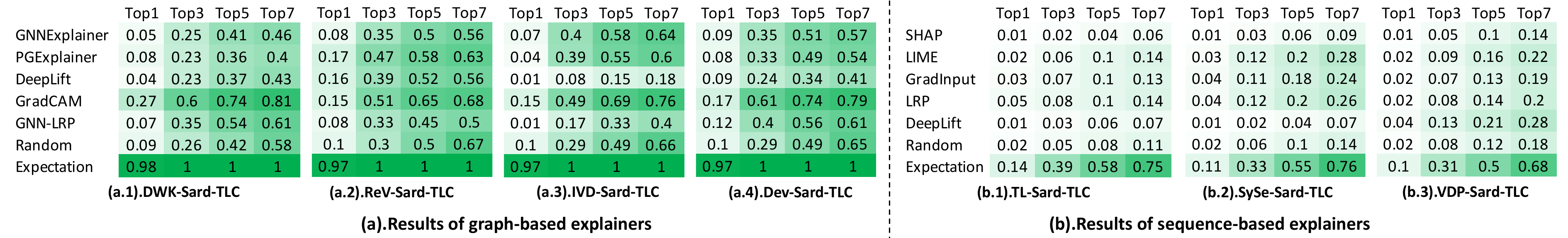}
    \vspace{-4mm}
    \caption{Mean TLC of graph-based and sequence-based explainers on SARD datasets.
\label{experiment:trigger_coverage}}   
\end{figure}

Fig.~\ref{experiment:trigger_coverage} (a) presents the average TLC results for the five graph-based approaches. It allows us to observe the following patterns:

\par \textbf{\textit{None of these graph-based explainers attain satisfactory TLC  scores.}} We can observe that nearly all methods exhibit a significant deviation from Expectation. While the best results of GradCAM is only about 80\% when explaining DeepWuKong as shown in Fig.~\ref{experiment:trigger_coverage} (a.1), it also produces worse results on Reveal as illustrated in Fig.~\ref{experiment:trigger_coverage} (a.2).
As for the other four explainers, their performance is worse than Random in most cases.

\par \textbf{\textit{Analyzing TLC results alongside fidelity scores reveals a weak correlation between the two.}}
For instance, DeepLift achieves the highest fidelity+ scores when explaining DeepWuKong and Reveal as shown in Fig.~\ref{experiment:fidelity_graph} (a.2),(b.2).
Yet its TLC is lower than Random for these detectors. 
Notably, when explaining IVDetect, DeepLift's TLC scores are significantly lower than Random as illustrated in Fig.~\ref{experiment:trigger_coverage} (a.3). 
On the other hand, GradCAM performs best in terms of TLC among the five approaches. Still, its fidelity+ scores are notably lower than DeepLift's when explaining the four detectors as depicted in Fig.~\ref{experiment:fidelity_graph} (a.2),(b.2),(c.2),(d.2).

\subsubsection{FLC Among Graph-based Explainers}\hfill

\begin{figure}[h!]
  \includegraphics[width=1.0\textwidth]{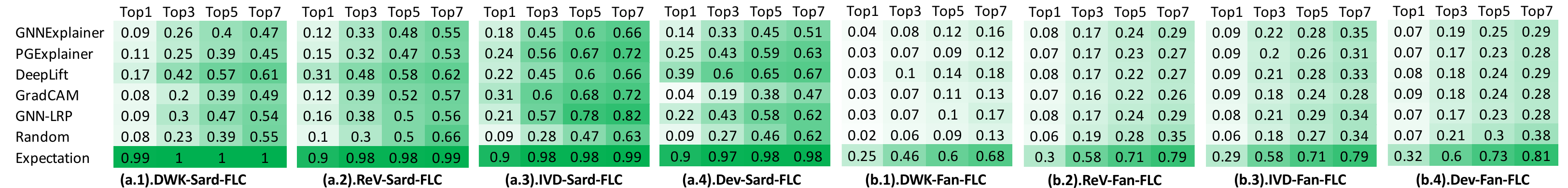}
    \caption{Performance of graph-based explainers in terms of FLC.
\label{experiment:graph_fix_coverage}}   
\end{figure}

Fig.~\ref{experiment:graph_fix_coverage} displays the experimental results of FLC for the five graph-based explainers on both the SARD and Fan datasets. 
The following observations can be made:

\par \textbf{\textit{The FLC results of graph-based explainers on both datasets fall below expectations.}} On the Fan dataset, 
the FLC scores of the five explanation approaches on Reveal and Devign are notably lower than Random as shown in Fig.~\ref{experiment:graph_fix_coverage} (b.2),(b.4), not to mention the comparison with Expectation.
When explaining DeepWuKong and IVDetect, DeepLift and GNNExplainer perform the best over other methods respectively.
However, their performance still falls considerably short, being even worse than half of Expectation as shown in Fig.~\ref{experiment:graph_fix_coverage} (b.1),(b.3).
On the SARD dataset, DeepLift attains the best FLC scores when explaining DeepWuKong, Reveal, and Devign. Nevertheless, all its top-7 FLC are below 70\%.
As for GNN-LRP, although it achieves the highest top-7 FLC of 82\% when explaining IVDetect as shown in Fig.~\ref{experiment:graph_fix_coverage} (a.3), there is a significant decrease in performance when explaining the other three detectors.

\par \textbf{\textit{The results of the explanation methods on FLC show little correlation to TLC and fidelity.}} On both SARD and Fan dataset, the fidelity+ scores and FLC scores do not align. 
Despite GNNExplainer achieving the highest fidelity+ scores when explaining IVDetect shown in Fig.~\ref{experiment:fidelity_graph} (c.2), its FLC results are worse than PGExplainer, GradCAM, and GNN-LRP in this case depicted in Fig.~\ref{experiment:graph_fix_coverage} (a.3). 
On the Fan dataset, when explaining IVDetect, GNNExplainer performs 1\%-4\% better than Random as depicted in Fig.~\ref{experiment:graph_fix_coverage} (b.3).
However, there is a significant difference in fidelity+ scores between Random and the other methods as shown in Fig.~\ref{experiment:fidelity_graph} (c.4), indicating a lack of correlation between FLC and fidelity+.
Similarly, the FLC scores do not correspond to the TLC scores. While GradCAM achieves the highest TLC scores on DeepWuKong as illustrated in Fig.~\ref{experiment:trigger_coverage} (a.1), its corresponding FLC scores are significantly lower than those of DeepLift and GNN-LRP as shown in Fig.~\ref{experiment:graph_fix_coverage} (a.1).


\begin{figure*}[t]
  \hspace{0mm}
  \vspace{-1mm}
  \includegraphics[width=1.0\textwidth]{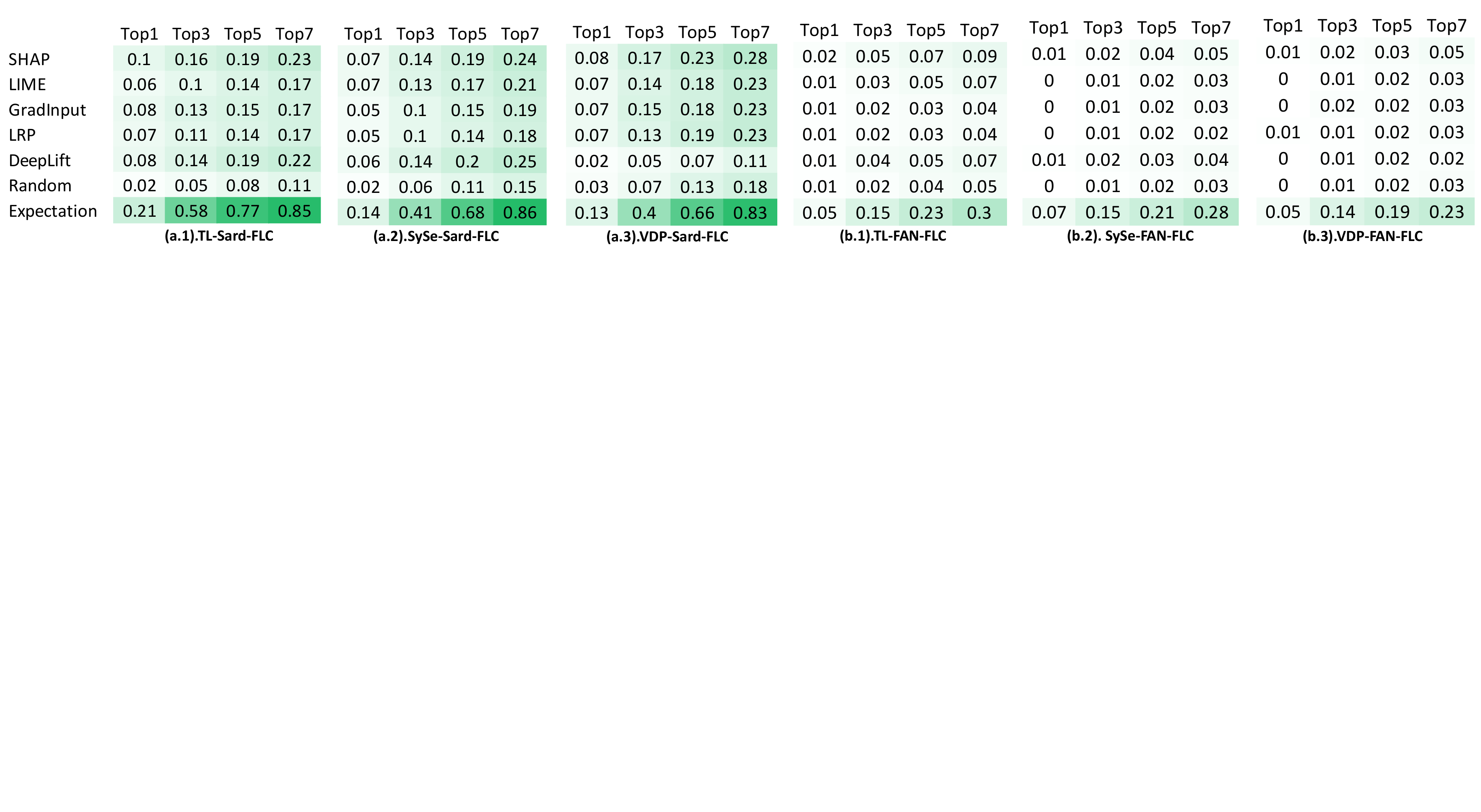}
    \vspace{-58mm}
    \caption{Performance of sequence-based explainers in terms of FLC.
\label{experiment:sequence_fix_coverage}}
\vspace{-5mm}
\end{figure*}

\subsubsection{TLC Among Sequence-based Explainers}\hfill

Fig.~\ref{experiment:trigger_coverage} (b) presents the TLC scores of sequence-based explainers, indicating:

\par \textbf{\textit{Sequence-based explainers obtain unsatisfactory TLC scores.}} 
Similar to the situation in graph-based explainers, the TLC scores of the sequence-based explainer significantly deviate from the Expectation.
When explaining SySeVR, Lime produces the highest TLC scores, which are nearly twice than Random as shown in Fig.~\ref{experiment:trigger_coverage} (b.2), however, the scores are even lower than half of the Expectation.
The same observation holds true for the other two detectors.

\par \textbf{\textit{The fidelity scores of sequence-based explainers exhibit a limited correlation with TLC scores.}} Analyzing the results in terms of fidelity scores, we observe that higher fidelity scores do not necessarily correspond to higher TLC.
Taking Lime for example, it attains the highest fidelity scores among five sequence-based explainers as shown in Fig.~\ref{experiment:fidelity_seq} (c.1). However, in regard to TLC scores, it is inferior to DeepLift on VulDeePecker as depicted in Fig.~\ref{experiment:trigger_coverage} (b.3), while DeepLift produces significantly lower fidelity scores than Lime.

\subsubsection{FLC Among Sequence-based Explainers}\hfill

Fig.~\ref{experiment:sequence_fix_coverage} illustrates the FLC results of sequence-based explainers, it reveals that:

\par \textbf{\textit{The FLC scores achieved by sequence-based explainers fall short of expectations.}} On the Fan dataset, when explaining TokenLSTM, Shap attains the best FLC score, while it only reaches 30\% of Expectation as shown in Fig.~\ref{experiment:sequence_fix_coverage} (b.1).
On the SARD dataset, Shap performs the best in FLC among the five explainers, however, the observed outcome diverges significantly from Expectation as depicted own in Fig.~\ref{experiment:sequence_fix_coverage} (a.1),(a.2),(a.3).


\par \textbf{\textit{There is minimal correspondence between the fidelity scores and FLC.}} One example is that Shap continues to outperform other explainers in terms of FLC on the SARD dataset. However, it does not achieve the highest fidelity scores across all detectors as shown in Fig.~\ref{experiment:fidelity_seq} (a.1),(b.1),(c.1).

Considering the aforementioned observations, it is uncertain whether explanation methods can reliably provide valuable insights. 
We hypothesize that DL-based detectors aim to learn the distinctions between vulnerable and normal samples, attempting to discern whether the content represents a vulnerability or normal code fragments. 
Given that VTSs exist in both vulnerable and non-vulnerable programs, DL-based detectors might not prioritize the importance of VTS in determining vulnerabilities. 
Hence, theoretically, explanation methods should achieve high FLC. 
As they focus on the differences between vulnerable and normal samples. 
However, experimental outcomes do not consistently align with this expectation. 
Furthermore, there appears to be no direct correlation observed among fidelity, TLC, and FLC values. 
As suggested by T  Ganz~\cite{ACMExplainDiscovery}, this discrepancy might stem from DL-based detectors being trained not to recognize vulnerabilities but artifacts encompassing not only VFS. 
These findings underscore the need for further analysis of DL-based detectors and explanation approaches, which will be discussed in \S~\ref{RQ3}.




\vspace{2mm}
\noindent\fbox{
	\parbox{0.95\linewidth}{
ANSWER: The LC scores of these explanations indicate their failure to provide sufficient vulnerability context in some cases, rendering them unsuitable for working with DL-based detectors in identifying vulnerability-related statements. 

}}
\vspace{2mm}

\subsection{RQ3: Reliability of DL-based detectors}\label{RQ3}

\baijun{In this section, we investigate whether DL-based detectors have learned features unrelated to vulnerabilities, leading to unreliable explanation results. 
This research comprises two parts. In the first segment~(RQ 3.1), we initially sample a sub-dataset from the original full dataset, ensuring an equal quantity of vulnerable and non-vulnerable samples while controlling the differences between vulnerable and normal samples primarily within VFS. 
This sub-dataset is partitioned into train, evaluation, and test sets. 
We retrain a model, comparing the detection performance of the retrained model with the previous model on the sub-test dataset. 
When the performance difference between the retrained and previous models is minimal, we compare the performance differences of explanation methods between the retrained and previous models. Additionally, we introduce two metrics to measure the sensitivity of the retrained and previous models to VTS and VFS, as well as their susceptibility to disturbances irrelevant to vulnerabilities.
In the second part~(RQ 3.2), we aim to evaluate the detectors’ capacity to identify previously unseen vulnerability-fixing patterns. 
This assessment intends to discern whether DL-based detectors genuinely comprehend vulnerabilities or merely recognize artifacts.}

\subsubsection{RQ3.1: Comparsion between previous and retrained model}\label{sec:cmp}\hfill

\paragraph{Experimental Setting}

\baijun{In this part, we aim to investigate the impact of DL models themselves on the explanation methods' ability to pinpoint vulnerable statements. 
We approach this by first sampling a sub-dataset from the full dataset, ensuring pairs of vulnerable and normal samples while emphasizing differences primarily within the vulnerability-fixing lines. 
Subsequently, we divide this sub-dataset into three segments—train, eval, and test—following an 8:1:1 ratio. 
Based on this sub-dataset, we retrain a model, ensuring similar detection performance between the retrained and previous models. 
Under this condition, we experiment with explanation methods. 
To measure their performance, we employ three metrics: fidelity, TLC, and FLC. For graph-based explainer, we utilize fidelity+ as a supplemental fidelity metric. 
Theoretically, the retrained model, due to the more pronounced differences between vulnerable and normal samples in the training set, should display increased sensitivity to VFS after retraining. 
Consequently, this heightened sensitivity should enhance the explanation methods' ability to pinpoint parts related to vulnerabilities.}

\baijun{\textbf{Dataset and detector consideration.} When sampling the dataset, we chose to sample from the SARD portion and not from the Fan dataset. 
This decision was influenced by concerns regarding the potential presence of vulnerability-unrelated code changes in patch commits, as discussed in~\cite{dataQuality}. 
Such changes could result in inaccurate labeling of vulnerabilities within the dataset.
Simultaneously, we solely employ function-level detectors for experimentation and refrain from using slice-level detectors. 
This decision is based on the fact that DeepWuKong, SySeVR, and VulDeePecker detectors conduct slicing and symbolization operation. 
This process potentially exacerbates the differences unrelated to vulnerabilities between vulnerable and normal samples. 
For example, during symbolization, variable names such as ``data'' or ``databuffer'' in the original samples may be symbolized as ``VAR1'' and ``VAR2''. However, in different samples or between a vulnerability sample and its corresponding normal sample, these variables may be symbolized as different values, such as ``VAR1'' in the vulnerability sample and ``VAR2'' in the normal sample. 
Subsequent experiments~(\S~\ref{sec:turnover}) reveal that these textual disparities interfere to some extent with DL-based detectors.}

\baijun{\textbf{Balanced sub-dataset curation.} For the samples within the full SARD dataset, we collect each vulnerable function and its corresponding patched version based on the test case ID. 
Subsequently, we extract samples with the same number of lines in both the vulnerable and normal functions, forming a new sub-dataset. 
This process entails some lines within the vulnerable functions being replaced by others in the patched version. 
This method is employed considering our observation: within SARD, vulnerable functions and their corresponding patched functions, with an identical number of lines, generally exhibit similar graph structures or possess token sequences with comparable lengths. 
This practice helps to minimize differences related to vulnerability-unrelated aspects between the vulnerable and normal samples as much as possible.
}

\begin{figure}[t]
  \centering
    \includegraphics[width=0.8\textwidth]{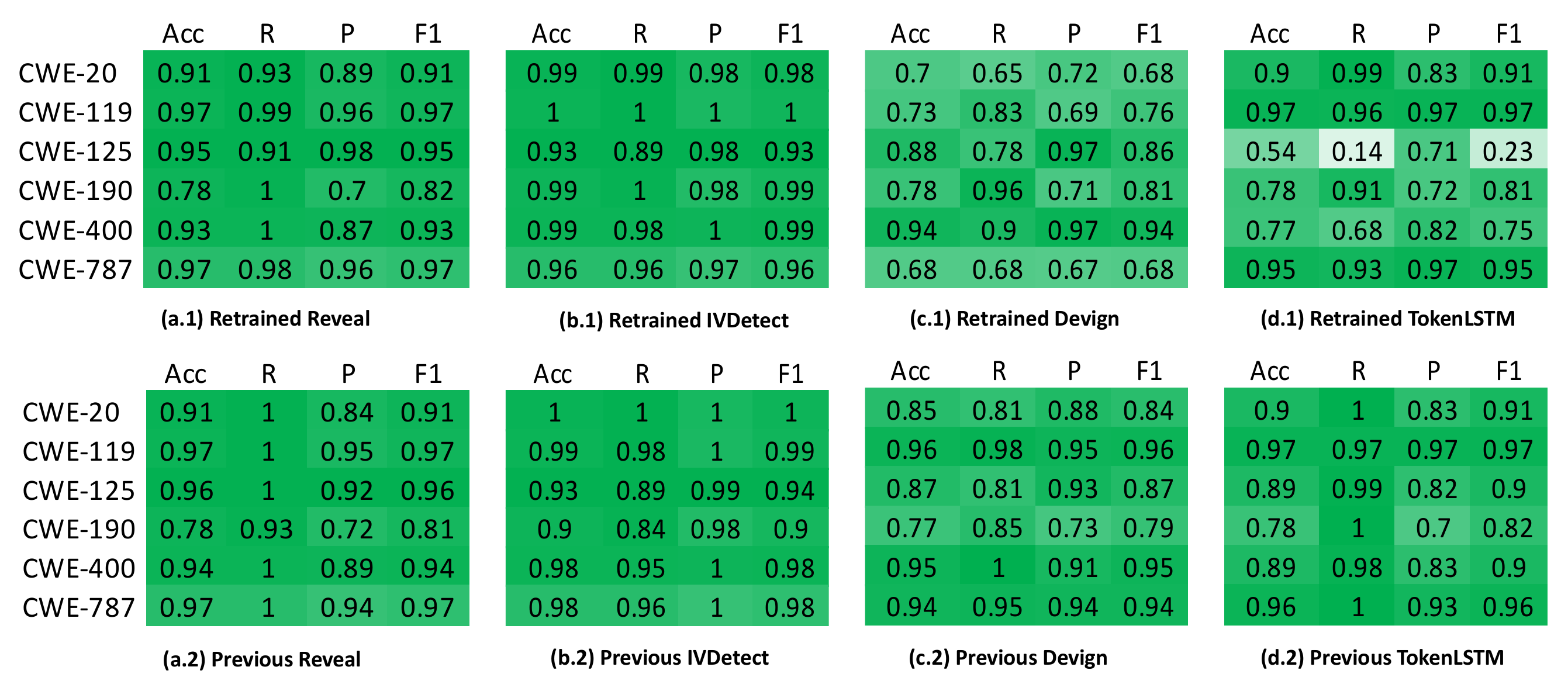}
  \vspace{-3mm}
  \caption{Detection performance of four function-level detectors with retrained and previous models on sub-test-dataset.}
  \vspace{-5mm}
  \label{fig:retrain}
\end{figure}

\baijun{\textbf{Dataset and detector selection.} After sampling, the remaining vulnerable sample sizes in the sub-test set are shown in Table~\ref{tab:num_of_sample}, indicating a considerable reduction in the dataset sizes for each type of vulnerability. The most notable reductions are observed in CWE-190 and CWE-400. 
On the sub-dataset, we first retrain the models for Reveal, IVDetect, Devign, and TokenLSTM. 
The comparison of the performance on the sub-test dataset between the previous and retrained models for each detector is depicted in Fig.~\ref{fig:retrain}.
We can observe that Reveal and IVDetect exhibit minor performance differences between the previous and retrained models across the six types of sub-test datasets. 
However, Devign shows performance declines in CWE-20, CWE-119, and CWE-787, while TokenLSTM specifically demonstrates reduced performance in CWE-125, CWE-190, and CWE-400. 
We speculate that the performance declines may be related to the quantity and quality of the training samples. 
Considering the requirement for similar performance between the previous and retrained models, the number of test samples, and the need to involve both graph-based and sequence-based methods, we opt to continue the experiments on portions of CWE-20 and CWE-119 samples for Reveal, IVDetect, and TokenLSTM.}

\begin{table}[t]\Large
  \caption{Number of vulnerable functions in each testset after filtering.}
  \begin{adjustbox}{width=0.65\textwidth,center}
      \centering
      \begin{tabular}{c|c|c|c|c|c} 
      \hline
      CWE-20 & CWE-119 & CWE-125 & CWE-190 & CWE-400 & CWE-787 \\
      \hline
      804 & 875 & 172 & 91 & 41 & 664 
      \\
      \hline
      
    \end{tabular}
  \end{adjustbox}
  \label{tab:num_of_sample}
 \vspace{-7mm}
\end{table}

\baijun{\textbf{VUR metric.} To explore the susceptibility of DL-based detectors to perturbations unrelated to vulnerabilities, we introduce a novel metric termed VUR (Vulnerability Unrelated Ratio). VUR quantifies the model's susceptibility to perturbations unrelated to vulnerabilities and is determined by the following equation:}

\begin{equation}\label{equa:vur}
   VUR = \frac{\sum_{sc_i}(\hat{f}(x) - \hat{f}(x^{-sc_i})) . (|sc_i \cap (VFS \cup VTS)| = 0)}{\sum_{sc_i}(\hat{f}(x) - \hat{f}(x^{-sc_i}))}. 
\end{equation}

\baijun{Here, $\hat{f}(x) = (f(x) > 0.5)$ represents the predicted label of the corresponding detector $f$, and $sc_i$ denotes a set of nodes or tokens associated with a perturbation. 
We exclusively calculate VUR on TP samples.
Thus, $\hat{f}(x) = 1$. 
When $\hat{f}(x^{-sc_i}) = 0$, it signifies that the perturbation $sc_i$ has successfully flipped the model's prediction. 
Hence, the denominator in Equation~\ref{equa:vur} denotes the number of perturbations within a given perturbation set capable of flipping the model's prediction. 
The term $(|sc_i \cap (VFS \cup VFS)| = 0)$ signifies whether $sc_i$ has no intersection with VFS or VTS, which equates to 1 if there is no intersection. 
Therefore, the numerator in Equation \ref{equa:vur} computes the count of perturbations within the perturbation set capable of flipping the model's prediction and having no intersection with VFS or VTS.
VUR signifies the proportion of perturbations capable of flipping the model's predicted label, which have no intersection with VTS and VFS. 
A higher VUR value indicates a greater susceptibility of the model to be influenced by perturbations that are unrelated to vulnerabilities.
Computing the VUR for all potential perturbations for each sample entails a significant computational cost. Therefore, in our computations, we randomly select 100 perturbations, each ranging in length from 1 to 3, for every sample to calculate the respective VUR. 
As a result, the VUR values for the retrained and previous models are not directly comparable. 
Instead, they indicate the susceptibility of different models to disturbances irrelevant to vulnerabilities.
In this context, VUR serves as a valuable metric for determining the suitability of perturbation-based evaluation metrics and explanation approaches in the context of vulnerability detection. 
An elevated VUR score suggests that higher-fidelity explanations may not necessarily convey vulnerability-related information.}

\baijun{To further probe the importance of the VTS and VFS for DL-based detectors. 
We define a metric called $imp$ which is similar to fidelity. 
It could be denoted as $imp_{VTS/VFS} = f(x) - f(x^{-VTS/VFS})$, where $imp_{VTS/VFS}$ indicates the importance score for VTS or VFS, $x$ is an input code fragment, $x^{-VTS/VFS}$ is the sample after masking VTS or VFS.}

\paragraph{Experimental Results}

\begin{figure}[t]
  \centering
    \includegraphics[width=0.9\textwidth]{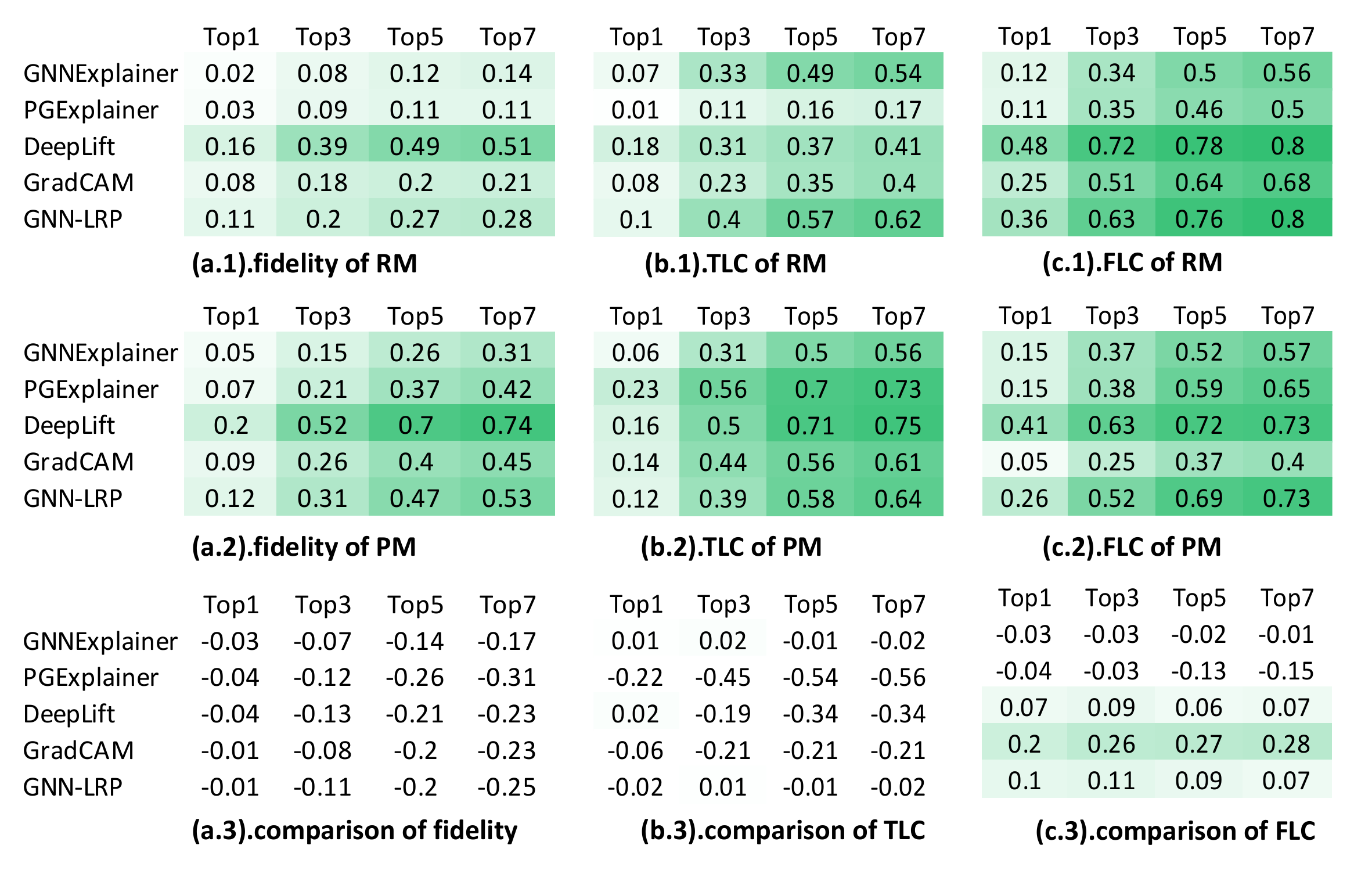}
  \vspace{-3mm}
  \caption{Explanation results for Reveal. ``RM'' denotes ``retrained model'', ``PM'' denotes ``previous model'', ``comparison'' denotes the performance comparison between retrained and previous models.}
  \vspace{-5mm}
  \label{fig:cmp_reveal}
\end{figure}

\begin{figure}[t]
  \centering
    \includegraphics[width=0.9\textwidth]{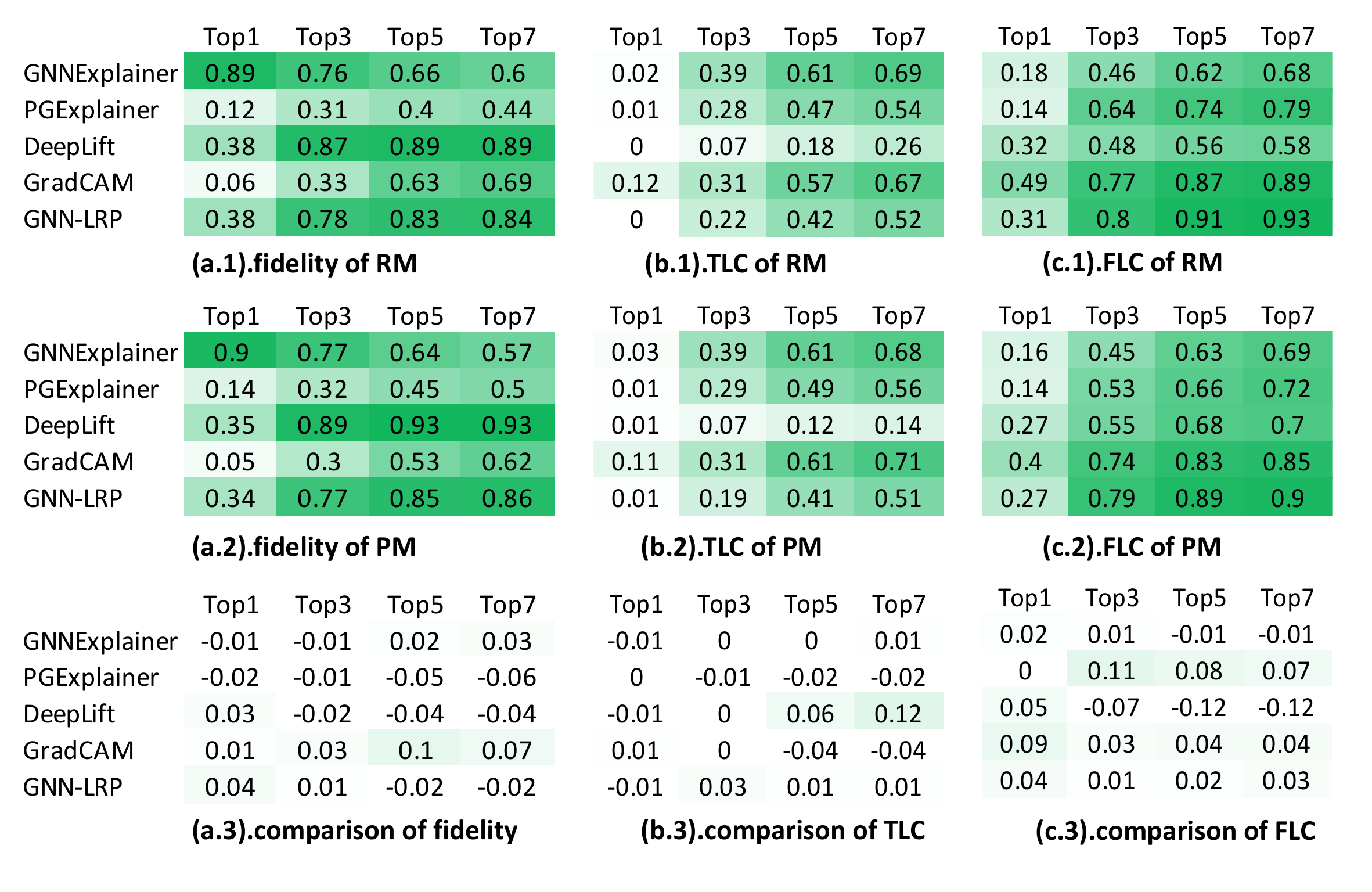}
  \vspace{-3mm}
  \caption{Explanation results for IVDetect.}
  \vspace{-5mm}
  \label{fig:cmp_ivdetect}
\end{figure}

\begin{figure}[t]
  \centering
    \includegraphics[width=0.9\textwidth]{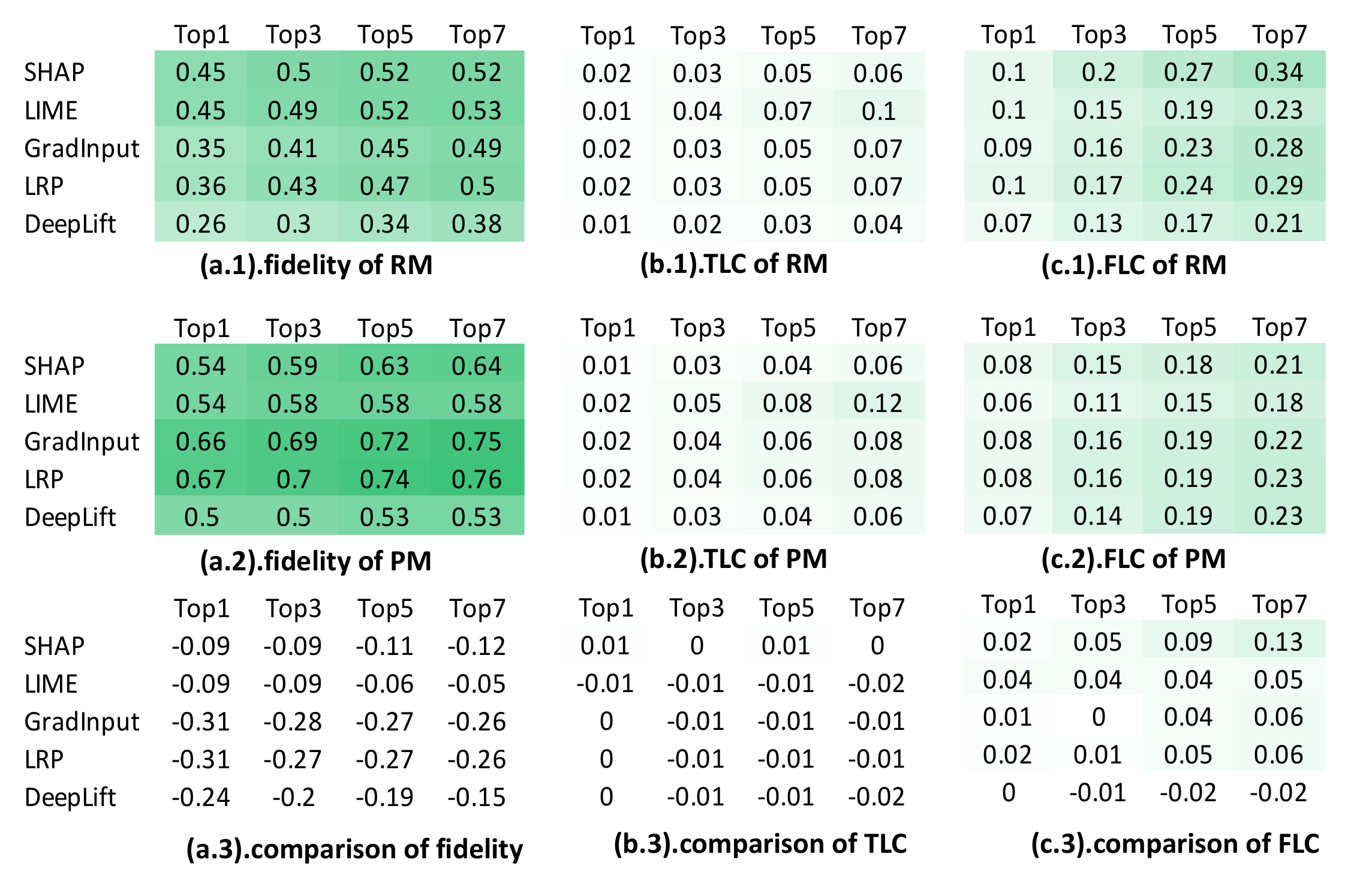}
  \vspace{-3mm}
  \caption{Explanation results for TokenLSTM.}
  \vspace{-5mm}
  \label{fig:cmp_tokenlstm}
\end{figure}

\baijun{The explanation results for Reveal, IVDetect, and TokenLSTM are depicted in Fig.~\ref{fig:cmp_reveal},~\ref{fig:cmp_ivdetect} and ~\ref{fig:cmp_tokenlstm} respectively. From the results, we can observe: }

\baijun{\textbf{Firstly, perturbation-based evaluation metrics or explanation approaches may not be suitable for vulnerability detection.} 
It is evident that the level of fidelity does not inherently correlate with FLC and TLC. 
In the majority of explanation methods, despite achieving higher FLC on the retrained model compared to the previous model, their fidelity values tend to decrease. 
Specifically in the context of explaining IVDetect, both DeepLift and GNNLRP exhibited respectable fidelity on the previous model.
However, GNNLRP displayed higher TLC and FLC compared to DeepLift.
Therefore, fidelity might not be a reliable metric as disturbances irrelevant to vulnerabilities could significantly impact the model's predictions.}

\baijun{The VUR results for retrained and previous detectors are presented in Table~\ref{tab:vur}. 
Notably, in both models, a significant proportion of perturbations unrelated to vulnerabilities have the capacity to invert the detectors' labels. 
We perform an assessment of all detectors using the previous models on the full-test dataset, and the outcomes are detailed in Table~\ref{tab:vur_full}. 
Remarkably, this assessment reiterates that vulnerability-unrelated perturbations possess the capability to flip the predictions of all detectors.
Consequently, explanations with high fidelity might not necessarily be correlated with vulnerabilities. Therefore, using fidelity as an evaluation metric in the context of vulnerability detection might not be suitable. 
Additionally, existing perturbation-based methods like GNNExplainer, PGExplainer, etc., which aim to maximize the impact on model prediction metrics, are also unsuitable for deployment in the realm of vulnerability detection.
}


\textbf{DL-based detectors are focused on differentiating vulnerable and normal samples, a characteristic also mirrored by gradient-, decomposition-, and surrogate-based explainers.} 
Within Reveal, all model-related explainers demonstrate an ability to achieve higher FLC scores while concurrently registering decreased TLC.
Regarding IVDetect, while the FLC score of DeepLift remains relatively modest without notable improvements, both GradCAM and GNN-LRP consistently maintain notably high FLC scores, exhibiting a slight increase, particularly evident in the retrained model.
In the case of TokenLSTM, it is notable that all deployed explainers, with the exception of DeepLift, attain higher FLC scores in the retrained models.
The aforementioned observations suggest that the retrained model indeed tends to place more emphasis on the VFS during predictions compared to the previous model. 
Additionally, explanation methods, beyond perturbation-based approaches, to some extent, can capture this particular aspect of the model.
The results depicted in Table~\ref{tab:imp} reveal that in both the retrained and previous models, the significance of VFS surpasses that of VTS.
Nevertheless, when examining their explanations in both retrained and previous models, their results may still not be considered ideal. 
This implies that they are still prone to selecting features unrelated to vulnerabilities, which could be associated with both the deep learning models and the explanation methods themselves. 
However, discussing this aspect is beyond the scope of our study.

\baijun{For these three detectors, the $imp_{VTS}$ in the retrained model has decreased, aligning with our hypothesis that ``VTS exist in both vulnerable and normal samples, leading the detectors to not assign high importance during prediction.''
However, concerning $imp_{VFS}$, in the case of IVDetect and TokenLSTM, the scores have improved after retraining. In contrast, Reveal shows a slight decline after retraining, deviating from our expectations. 
Yet, considering the earlier discussion, perturbation-based evaluation metrics might not robustly quantify the significance of VFS and VTS for the model. 
Moreover, given the heightened FLC scores of the model-related explainers—DeepLift, GradCAM, and GNN-LRP in Reveal, it consistently indicates the model's inclination toward assigning greater importance to VFS during predictions on the sub-test dataset. 
Notably, these model-related explainers effectively identify and highlight this particular aspect.
However, the decrease in fidelity in the retrained model indicates that even if the model captures this fact, it might not necessarily be quantifiable through perturbation-based metrics.
}

\noindent\fbox{
	\parbox{0.95\linewidth}{
ANSWER: \baijun{DL-based detectors are susceptible to perturbations unrelated to vulnerabilities, resulting in the inversion of predicted labels. 
Hence, fidelity, a perturbation-based evaluation metric, is not suitable for assessing the performance of explanation methods in the context of vulnerability detection. 
Additionally, existing perturbation-based explainers are not applicable in this scenario.
Furthermore, explainers beyond perturbation-based ones have the capacity to capture information related to the models' prediction process.
Nonetheless, all explanation approaches are still susceptible to selecting features unrelated to vulnerabilities.}
}}

\begin{table}[t]\Large
  \caption{VUR results of detectors. ``R'' denotes retrained model, ``P'' demotes previous model.}
  \begin{adjustbox}{width=0.4\textwidth,center}
      \centering
      \begin{tabular}{c|c|c|c|c|c} 
      \hline
      R-ReV & P-ReV & R-IVD & P-IVD  & R-TL & P-TL \\
      \hline
      0.47 & 0.52 & 0.48 & 0.26 & 0.36 & 0.37 
      \\
      \hline
    \end{tabular}
  \end{adjustbox}
  \label{tab:vur}
 \vspace{-3mm}
\end{table}

\begin{table}[t]\Large
  \caption{VUR results of detectors with previous models on full datasets.}
  \begin{adjustbox}{width=0.42\textwidth,center}
      \centering
      \begin{tabular}{c|c|c|c|c|c|c} 
      \hline
      ReV & IVD & Dev & DWK  & SySe & VDP & TL \\
      \hline
      0.78 & 0.90 & 0.57 & 0.71 & 0.66 & 0.43 & 0.78 
      \\
      \hline
    \end{tabular}
  \end{adjustbox}
  \label{tab:vur_full}
 \vspace{-3mm}
\end{table}

\begin{table}[t]\Large
  \caption{$imp_{VTS}$ and $imp_{VFS}$ for retrained and previous detectors. Higher $imp$ scores reflect the significance of VTS/VFS in model decision-making.}
  \begin{adjustbox}{width=0.5\textwidth,center}
      \centering
      \begin{tabular}{c|c|c|c|c|c|c} 
      \hline
      Detector & R-ReV & P-ReV & R-IVD & P-IVD  & R-TL & P-TL \\
      \hline
      $imp_{VTS}$ & 0.04 & 0.094 & 0.001 & 0.03 & 0.114 & 0.14 
      \\
      \hline
      $imp_{VFS}$ & 0.154 & 0.179 & 0.297 & 0.214 & 0.59 & 0.527
      \\
      \hline
    \end{tabular}
  \end{adjustbox}
  \label{tab:imp}
 \vspace{-3mm}
\end{table}

\subsubsection{RQ3.2: Probing DL-based detectors}\label{sec:turnover}\hfill

\begin{figure}[t]
  \centering
    \includegraphics[width=0.5\textwidth]{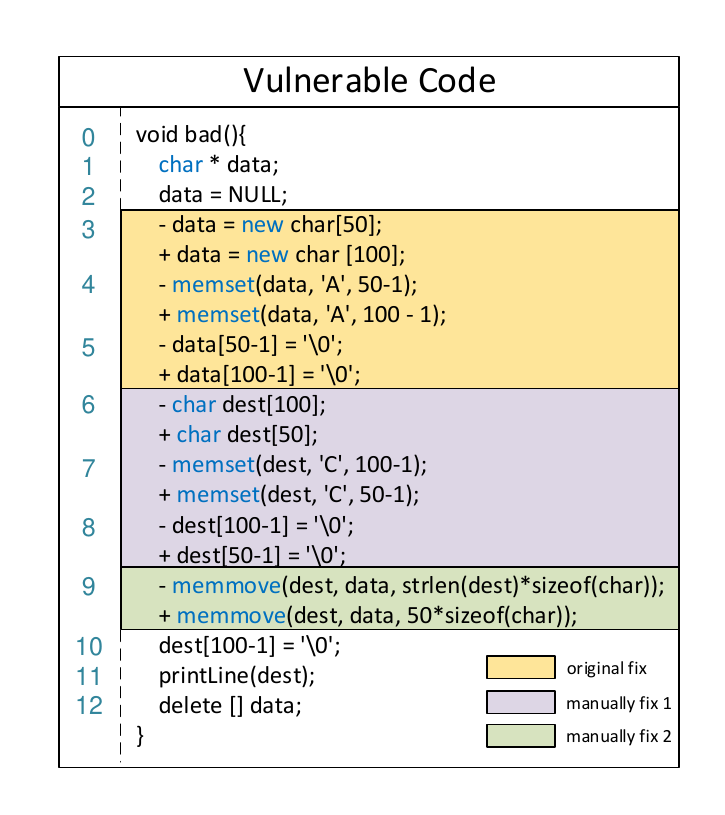}
  \vspace{-3mm}
  \caption{Examples of the original fixed version and manually fixed version.}
  \vspace{-5mm}
  \label{fig:fixCode}
\end{figure}

\baijun{In our previous experiments, we observed that existing DL-based detectors are susceptible to disturbances unrelated to vulnerabilities. 
Moreover, they exhibit a clearer distinction between vulnerability and normal samples, relying predominantly on VFS for predictions, particularly if there are almost no discrepancies present in the training set apart from the VFS.
However, what remains unclear is whether DL-based detectors merely discern differences between positive and negative samples to identify vulnerable code, or if they genuinely comprehend the mechanisms triggering vulnerabilities. 
In this part, our objective is to evaluate the detectors' capacity to recognize previously unseen vulnerability-fixing patterns. 
This analysis aims to provide further insights into this particular aspect.}

We manually selected 100 TP samples and their corresponding fixed versions from each sub-dataset of SARD. 
Various methods were employed to manually correct the vulnerable samples. 
For example, Fig.~\ref{fig:fixCode} showcases \baijun{a piece of patched vulnerable code example, ``-'' denoted removed lines from vulnerable code, ``+'' denotes added lines in the patched version.
Where lines 3-5 are originally fixed ways. lines 6-8 and 9 are two different manually fixed ways.} 
During the manual correction process, we ensured that all the tokens incorporated into the code were already present in the original SARD dataset to avoid the out-of-vocabulary~(OOV) problem of the embedding models. Subsequently, we evaluated both corrected code versions using vulnerability detectors and compared their respective accuracies.
Following experiments in \S~\ref{sec:cmp}, we opt not to carry out this experiment on the Fan dataset for the same reason.

The results are presented in Table~\ref{table:turnover_rate}, highlighting a significant decrease in accuracy for the detectors when evaluated on the manually corrected samples compared to the original corrected samples. The difference in results ranges from 55\% to 87\% across each detector. These findings suggest that the DL-based detectors are unable to recognize previously unseen fixing patterns, indicating a lack of effectiveness in capturing the underlying triggering mechanism of vulnerabilities.

\begin{figure}[t]
  \centering
    \includegraphics[width=0.65\textwidth]{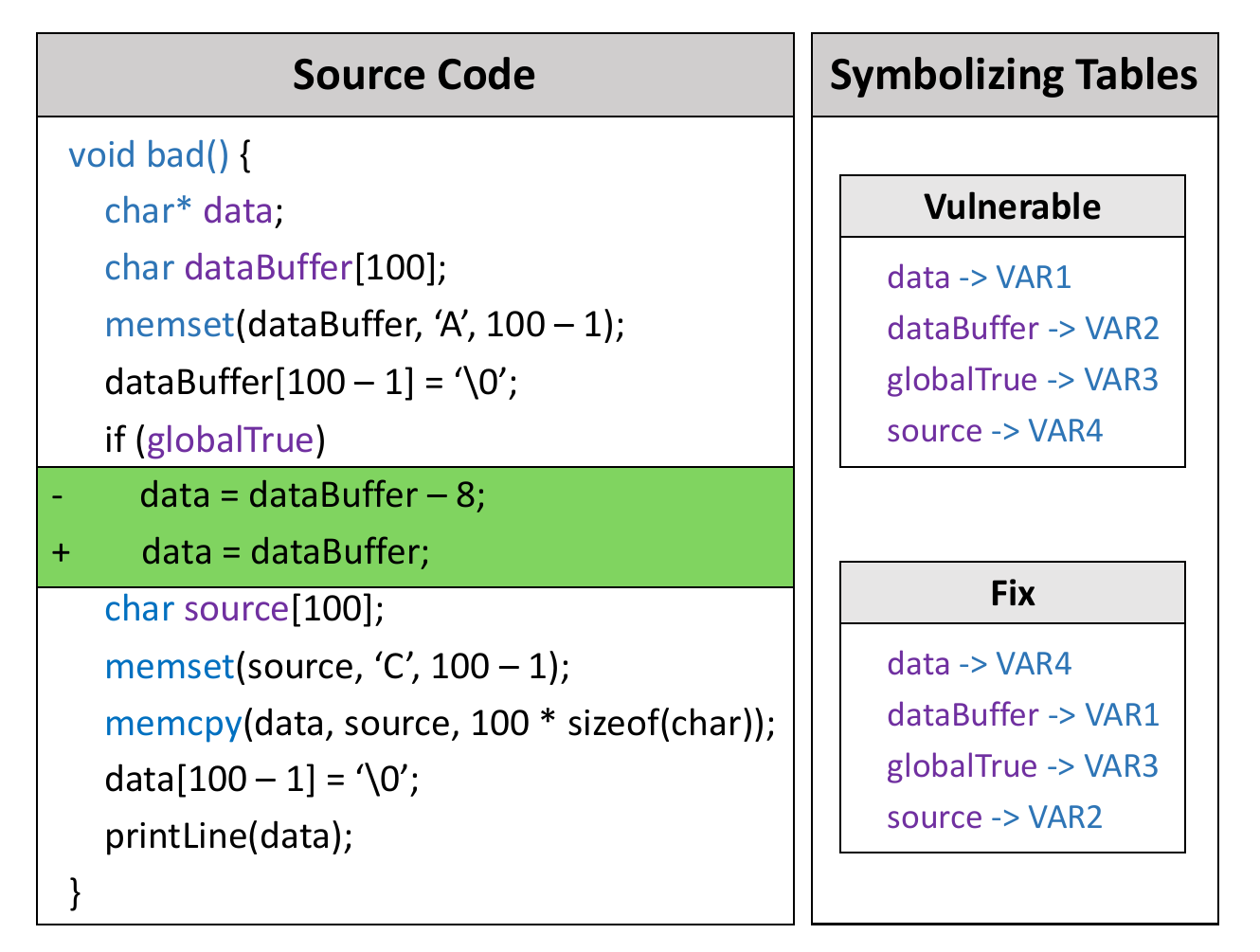}
  \vspace{-2mm}
  \caption{Examples illustrating problems of symbolizing.}
  \vspace{-2mm}
  \label{fig:symbolizing}
\end{figure}

Furthermore, during this experiment, we incidentally observe a text pattern capable of inducing misclassifications for DeepWuKong, as illustrated in Fig.~\ref{fig:symbolizing}.
It displays a vulnerable code snippet and its corresponding fixed version. 
During preprocessing for DeepWuKong, variable names are transformed into symbolic names, such as ``VAR'' followed by an index. 
However, issues in the preprocessing pipeline have resulted in different symbolic names being assigned to identical variables in the vulnerable and fixed codes. 
Surprisingly, DeepWuKong can still accurately predict the symbolized samples. 
To investigate further, we manually adjusted the symbolization by changing the variable order. 
Remarkably, DeepWuKong misclassifies both the vulnerable and fixed code in this scenario. 
This observation indicates that vulnerability detectors rely more on recognizing textual patterns rather than truly comprehending underlying vulnerability patterns. 
Consequently, this limitation may hinder the ability of explanation approaches to identify vulnerability lines.

\begin{table}[t]\Large
  \caption{Comparsion of accuracy between original fixed samples and manual fixed samples.}
  \begin{adjustbox}{width=0.5\textwidth,center}
      \centering
      \begin{tabular}{c|c|c|c|c|c|c|c} 
      \hline
      fix type & ReV & IVD & Dev & DWK & SySe & VDP & TL\\
      \hline
      original fix & 0.74 & 0.92 & 0.89 & 0.97 & 0.98 & 0.86 & 0.95 \\
      \hline
      manually fix & 0.19 & 0.32 & 0.20 & 0.10 & 0.19 & 0.14 & 0.11 \\
      \hline
    \end{tabular}
  \end{adjustbox}
  \label{table:turnover_rate}
  \vspace{-0.1in}
\end{table}

\vspace{2mm}
\noindent\fbox{
	\parbox{0.95\linewidth}{
ANSWER: \baijun{DL-based detectors primarily rely on differences learned from the training set between vulnerable and normal samples, notably emphasizing VFS to identify vulnerabilities rather than discerning the triggering mechanisms of vulnerabilities.
Also, DL-based detectors may learn vulnerability-unrelated features during training process.}
}}
\vspace{2mm}

%% file: relatedAndConclude.tex
\section{Discussions}\label{discussion}

\subsection{Limitations}

\textit{First, we are not rigorously studying whether these explanation methods descriptively and accurately explain the models' output.}
In our study, we attempted to employ perturbation-based methods to compute the fidelity of explanation results and assess the significance of VTS and VFS in determining the model's importance. 
However, subsequent observations revealed a paradoxical decrease in fidelity despite the retrained model showing higher FLC. 
Moreover, the model is indeed susceptible to perturbations that can alter the predicted labels, regardless of vulnerabilities. 
As a consequence, using perturbation-based assessments to determine the significance of certain lines of code for the model holds limited interpretive value. 
Therefore, we can only infer, through the comparison between retrained and previous models, that apart from perturbation-based techniques, other explanation methods can, to some extent, elucidate the model's reasoning process. 
However, it's important to note the inherent limitations of this evaluation approach.

\baijun{\textit{Second, evaluating the effectiveness of sequence-based explanation methods in locating vulnerabilities remains challenging}. We propose TLC and FLC metrics. 
However, sequence-based detectors provide token-level explanations, making it difficult to directly evaluate TLC and FLC. 
We approximate token recall from relevant code lines but face challenges due to granularity differences. 
Determining the most effective method for computing coverage rate in sequence-based explanations remains an open question.}

\textit{Third, Our research does not involve vulnerability detection methods based on large language models~ (LLMs).} The detectors we employed are primarily based on initial embeddings generated through methods like Word2Vec~\cite{word2vec}, followed by input into GNN or RNN for binary classification. 
However, with the introduction of LLMs such as CodeBert~\cite{codebert}, significant transformations have emerged in the methodology of code embeddings. 
These large language models often exhibit enhanced efficiency when performing code embeddings compared to traditional methods like Word2Vec.
Fu et al.~\cite{linevul} demonstrated that LLMs outperform IVDetect. However, we have identified limitations in current binary classification-based detectors, which primarily concentrate on distinctions between vulnerable and normal samples.


\subsection{Implications}

\textbf{Limitations of DL-based detectors.} During our research, we observe that DL-based detectors primarily learn distinguishing artifacts from the differences between vulnerable and normal samples in the training set to identify potential vulnerability segments, rather than comprehending the inherent semantic essence of vulnerabilities. 
This, in itself, poses significant challenges for DL-based vulnerability detection.

\begin{itemize}
    \item \textbf{Firstly, acquiring a high-quality vulnerability dataset is not straightforward.} We note that models retrained on a more refined sub-dataset exhibit a greater focus on the VFS compared to the previous models. 
    This indicates a relatively reduced susceptibility to artifacts unrelated to vulnerabilities. 
    However, acquiring such high-quality datasets in real-world scenarios poses considerable challenges, as highlighted by Croft et al.~\cite{dataQuality}, who discovered error rates in labeling vulnerabilities ranging from 20\% to 71\% within existing real-world datasets.

    \item \textbf{Secondly, facilitating the model to focus more on vulnerability-related lines during the training process is not a straightforward task.} Even when trained on an ideal dataset, models can still be susceptible to perturbations unrelated to vulnerabilities that could flip prediction results (which might also be associated with the nature of deep learning models). 
    Additionally, when encountering unseen fixing patterns or vulnerable patterns in the training set, models might still make misclassifications.
    This indicates that the current training methodologies may not necessarily yield an ideal detector.
\end{itemize}

\textbf{Insights for future research.} Therefore, performing vulnerability detection based on binary classification may not be an optimal or efficient approach. Note that Kai et al.~\cite{RepresentationLearning} found that while DL-based methods perform well in code classification and clone detection tasks, their performance in vulnerability detection is unsatisfactory. 
This indicates a lack of effective learning of vulnerability semantics by these detectors.
We suggest that future work consider dividing vulnerability detection into multiple sub-tasks rather than persisting with methods solely through binary classification tasks. 
For instance, they could follow the approach proposed by Sun et al.~\cite{gptscan}, which initially employs large language models~(LLMs) to identify variables and statements within the programs that correlate with vulnerabilities. 
Subsequently, they can use existing static analysis techniques to verify properties such as accessibility between these variables and statements, thereby identifying vulnerabilities. 
This approach allows vulnerability detection tools to leverage the latest advancements in deep learning for code comprehension, particularly the utilization of LLMs. 
Simultaneously, it mitigates the shortcomings in the reasoning abilities of deep learning models. 
Ultimately, this method optimizes the utilization of deep learning for enhanced and more efficient vulnerability detection.

\section{Related Work}\label{sec:related}

\subsection{Vulnerability Detection}

Malware detection and vulnerability detection are two key focus in software security.
Several rule-based detectors have been developed for identifying software vulnerabilities and malicious behavior, such as Clang Static Analyzer~\cite{CSA}, Infer~\cite{Infer}, Coverity~\cite{Coverity}, Fortify~\cite{Fortify}, Rats~\cite{RATS}, Checkmarx~\cite{Checkmarx}, SVF~\cite{SVF}, and MalWuKong~\cite{malwukong}. 
Typically, these detectors rely on human experts to define checking rules using traditional static analysis theories. However, this manual rule definition process can be time-consuming and labor-intensive.

Another research direction focuses on using deep learning (DL) algorithms to identify security vulnerabilities. 
These DL-based detectors~\cite{vdp, SySeVR, dwk, Russel, devign, ivdetect, reveal} can learn vulnerability patterns without the need for manually defining rules, which leads to improved accuracy compared to rule-based approaches.
These methods partition programs into multiple code fragments and train a DL model to predict whether those fragments contain vulnerabilities.
However, they can only identify vulnerable code fragments without locating detailed statements.

To address the limitation, researchers have proposed various approaches.
Li et al.~\cite{vuldeelocator} introduce the concept of granularity refinement and present BRNN-vdl, a model that can identify vulnerable tokens within input programs.
Fu et al.~\cite{linevul} employ a fine-tuned CodeBert~\cite{codebert} model to predict vulnerable functions and utilize its inner attention mechanism to identify critical sections.
Hin et al.~\cite{linevd} tackle vulnerability detection as a node-level classification problem, where they directly predict vulnerable statements.
These approaches aim to enhance the precision and accuracy of identifying specific vulnerability-related statements within code.
However, these approaches rely on detection models that require specific models to work, thus lacking generality.

\subsection{Explanation Approaches}

With the rapid advancements in deep learning, there has been a growing interest in developing explainability methods for black-box models to provide insights into the decision-making processes of those models.
\baijun{Therefore these explainability methods can be employed to assist vulnerability detectors in achieving more granular vulnerability detection.
Existing explainability techniques can typically be categorized into two distinct groups: instance-level methods and model-level methods. 
Model-level methods~\cite{olah2017feature,xgnn}  primarily focus on furnishing overarching insights and high-level comprehension for the explication of deep learning models. 
In particular, they investigate the correlations between input patterns and specific model behaviors. 
In contrast, instance-level methods are typically designed to provide explanations for specific inputs and can be further subcategorized into gradient-based methods, surrogate methods, perturbation-based methods, and decomposition methods.}

Surrogate methods~\cite{LIME, graphlime, pgmexplainer, relex} assume that the complexities in the neighboring regions of the input example are relatively low and can be effectively represented by a simpler surrogate model. They usually train interpretable models like random forests~\cite{RF}, to approximate the predictions of the black-box model. 
While perturbation-based methods~\cite{gnnexplainer, pgexplainer, SubgraphX, Zorro, CasualScreen, GraphMask, IDLP, VulExplainer} investigate how perturbations in input features impact the model's output score. 
By masking out trivial features and focusing on critical ones, these methods aim to identify the important features contributing to the models' decisions. 
Gradient-based methods~\cite{grad, gradcam++} typically compute the importance of input features by analyzing gradients or hidden feature maps.
Decomposition methods~\cite{LRP-LSTM, gnnlrp} assess the significance of input features by decomposing the initial model predictions into multiple terms.

\subsection{Explainable AI for SE}

\baijun{Despite significant advancements in deep learning that have greatly benefited automated software engineering, the lack of explainability continues to raise concerns in both the industry and the research community. For instance, questions may arise about why a particular code snippet is predicted to have a vulnerability or why specific comments are associated with a given piece of code. To address these issues, researchers are exploring the application of explainability methods within the field of software engineering.}

\baijun{In the field of software engineering, defect prediction stands out as one of the primary domains where explainability methods find extensive application. For instance, several defect prediction tools~\cite{defect1_wattanakriengkrai2020predicting, defect2_pornprasit2021jitline, defect3_zheng2021just} employ LIME~\cite{LIME} to interpret the predictions of deep learning models. To address the limitations of LIME, which relies on random perturbations and may yield suboptimal results, Chanathip et al.~\cite{pornprasit2021pyexplainer} have introduced PyExplainer to explain defect prediction models. Experimental results have proved its superiority over LIME.
Humphrey et al.~\cite{humphreys2019explainable}  have achieved defect prediction model interpretation by directly incorporating self-attention mechanisms.
In other subfields, Rabin et al.~\cite{rabin2021understanding} have developed a model-agnostic interpretability method, named Sivand, for method name prediction and variable name misuse prediction models. The key insight lies in the notion that by eliminating extraneous portions of input programs from the prediction process, one can gain a more profound understanding of the essential features involved in model inference.
Rabin~\cite{rabin2022syntax} applied the syntax-aware variant Sivand-Perses to the same code intelligence tasks. They observed that Sivand-Perses is not only faster but also produces more compact sets of critical tokens in the reduced programs. Furthermore, they found that these critical tokens could be employed in generating adversarial examples for as many as 65\% of the input programs.}

To evaluate the effectiveness of these explainers in the context of vulnerability detection, Ganz et al.~\cite{ACMExplainDiscovery} conducted a study using various explanation approaches for graph neural networks. They proposed criteria such as stability and descriptiveness to assess the explainers. The study revealed that all evaluated explanation methods exhibited deficiencies in at least two of the criteria.
Hu et al.~\cite{hu2023interpreters} extensively evaluated six popular interpreters on four graph-based vulnerability detectors. They found that these interpreters performed poorly in terms of effectiveness, stability, and robustness. Specifically, instance-independent methods outperformed others in effectiveness, perturbation-based methods showed greater stability, and instance-independent approaches provided more consistent results for similar vulnerabilities in terms of robustness.
Compared to the previous two works, our research uncovers additional concerns within DL-based detectors. 
Specifically, we identify that these detectors, despite achieving accurate detection performance, may still learn features unrelated to vulnerabilities. Moreover, they are susceptible to being misled by irrelevant perturbations, leading to a reversal in predictions, thereby influencing the effectiveness of explanation methods.

These explanation methods serve as valuable tools for gaining insights into the decision-making process of black-box models, and their application in vulnerability detection is an area of active research.

\section{Conclusion}\label{sec:conclusion}

In this study, we systematically evaluate the challenges in locating vulnerabilities with DL-based detectors and explanation approaches. 
Firstly, we evaluate the effectiveness of the explanation approaches using two metrics: fidelity and LC. We find that the results of fidelity and LC show little correlation. Meanwhile, these explanation methods show limited effectiveness in locating vulnerable lines. 
Secondly, subsequent experiments with DL-based detectors suggest that they might learn artifacts unrelated to the vulnerabilities, as retraining models on a balanced dataset can indeed improve model-related explainers' performance regarding FLC. This phenomenon may stem from vulnerability-unrelated distinctions between vulnerable and normal samples, thereby diminishing the efficacy of explanation approaches in identifying vulnerable lines.
Furthermore, even when trained on a balanced dataset, models remain susceptible to vulnerability-unrelated perturbations, undermining the reliability of fidelity calculations or explanation approaches based on perturbation.
Based on our findings, employing binary classification models directly for vulnerability detection may not be effective. 
We suggest that future work endeavors decompose vulnerability detection into multiple subtasks and utilize deep learning techniques to improve efficiency.
